\documentclass[aps,pre,superscriptaddress,showpacs,preprint,floatfix]{revtex4-1}

\usepackage{amssymb,mathrsfs,amsfonts}
\usepackage{amsmath}
\usepackage{epstopdf}
\usepackage{nicefrac}
\usepackage{listings}
\usepackage{upgreek}
\usepackage{physics}
\usepackage{multirow}
\usepackage{color}
\usepackage{textcomp}
\usepackage{bibnames}
\usepackage{appendix}
\usepackage{nicefrac}
\usepackage{listings}
\usepackage{upgreek}
\usepackage{physics}
\usepackage{multirow}
\usepackage{color}
\usepackage{natbib}
\usepackage{graphicx}
\usepackage{tabls}
\usepackage{afterpage}
\usepackage{amsthm}
\usepackage{lastpage}
\usepackage{empheq, etoolbox}
\usepackage{epstopdf}
\usepackage{changes}
\colorlet{Changes@Color}{red}  

\newcommand{\M}{{\cal M}}

\newcommand{\TaO}{{Ta$_2$O$_5$}\:}
\newcommand\p[2]{\frac{\partial #1}{\partial #2}}

\newcommand\ti[1]{\widetilde{#1}}

\usepackage{scalerel}
\usepackage{tikz}
\usetikzlibrary{svg.path}

\usepackage{orcidlink}

\begin{document}

\title
{Supersonic-subsonic transition region in radiative heat flow via self-similar solutions}
\author
{Elad Malka}
\email{elad.malka@mail.huji.ac.il}
\affiliation{Department of Physics, Nuclear Research Center-Negev, P.O. Box 9001, Beer Sheva 8419001, Israel}
\affiliation{Racah Institute of Physics, The Hebrew University, 9190401 Jerusalem, Israel}
\author
{Shay I. Heizler\,\orcidlink{0000-0002-9334-5993}}
\affiliation{Department of Physics, Nuclear Research Center-Negev, P.O. Box 9001, Beer Sheva 8419001, Israel}

\begin{abstract}

We study the radiative hydrodynamics flow of radiation-driven heat waves in hot dense plasmas, using approximate self-similar solutions. Specifically, we have focused on the intermediate regime between pure radiative supersonic flow and the pure subsonic regime. These two regimes were investigated both using exact self-similar solutions and numerical simulations, however, most of the study used numerical simulations, mainly because the radiative heat wave and the shock regions are not self-similar altogether. In a milestone work [J. Garnier et al., Phys. Plas., 13, 092703 (2006)], it was found that for a specific power law dependency temperature profile, a unique exact self-similar solution exists, that is valid for all physical regimes. In this work we approximate Garnier's exact solution for a general power-law temperature-dependency, using simple analytical considerations. This approximate solution yields a good agreement compared to numerical simulations for the different thermodynamic profiles within the expected range of validity. In addition, we offer an approximate solution for the energies absorbed in the matter, again, for a general power-law temperature profile. Our approximate self-similar solution for the energy yields very good results comparing to exact numerical simulations for both gold and $\mathrm{Ta_2O_5}$. We also set a comparison of our self-similar solutions with the results of an experiment for radiation temperature measurement in a hohlraum in low-density foams that is addressed directly, to the intermediate regime, yielding a good agreement and similar trends. The different models as well as the numerical simulations are powerful tools to analyze the supersonic-subsonic transition region.

\end{abstract}
\maketitle

\section{Introduction}
\label{intro}

Radiative heat transfer has a great importance for the physical description of many fluids dynamics problems~\cite{RMTV,Shestakov,Nath,menahem}. It plays an important role in the physical description of high energy density physics (HEDP) phenomena, in astrophysical systems~\cite{castor,drake} as in plasma laboratory experiments as well, usually conducted in high energy lasers facilities~\cite{lindl1995,rosen1996,lindl2004}. Radiative heat waves are an example of such phenomena, when the heat conduction mechanism in these high temperatures is via radiation heat conduction, rather than the electron heat conduction. The wave propagates mainly through absorption and black-body emission processes. These radiative heat waves are often called Marshak waves~\cite{Marshak,ps}.

The basic scenario of studying the dynamics of Marshak waves is a one dimension (1D) semi-infinite slab medium. The nature of the flow depends on the radiation source properties, as well as on the matter microscopic properties,~i.e.~the equation of state (EOS), and in particular the heat capacity, and the opacity. In this study we explore the radiative heat flow mainly using semi-analytic self-similar solutions. Although numerical simulations are now cheap and widely used, analytic solutions are still important and helpful in benchmarking numerical codes and their accuracy (recent fluids examples are~\cite{Ramsey,Ruby,menahem,guderly1,guderly2}). In this work, we assume that the radiation field is in local thermodynamic equilibrium (LTE) with the matter, meaning they are strongly coupled, and so the spectral distribution of the source is given by Planck's black body radiation law. Therefore, at any given time the incoming flux from a thermal radiation source can be characterized by a unique radiation temperature $T_s(t)$. The governing equations for describing the dynamics in Lagrangian form are~\cite{ps,HR}:
\begin{subequations}
	\label{full_eqs_simple}
	\begin{equation}
	\label{mass_s}
	\p{V}{t}-\p{u}{m}=0
	\end{equation}
	\begin{equation}
	\label{momentum_s}
	\p{u}{t}+\p{P}{m}=0
	\end{equation}
	\begin{equation}
	\label{energy_s}
	\p{e}{t}+P\p{V}{t}=\p{}{m}\left(\frac{c}{3\kappa_R}\p{\left(aT^4\right)}{m}\right)
	\end{equation}                
\end{subequations}
where $t$ is the time, $m(x,t)=\int_0^x \rho(x',t)dx'$ is the Lagrangian coordinate while $x$ is the Eulerian one, $V\equiv 1/\rho$ is the specific volume ($\rho$ is the mass density), $u$ is the matter velocity, $P$ is the pressure, $e$ is the matter specific internal energy per mass unit, $T$ is the temperature and $c$ is the speed of light. $\kappa_R$ is the Rosseland mean opacity and $a\equiv 4\sigma/c$ is the radiation constant, where $\sigma$ is the Stefan-Boltzmann constant. At time $t=0$ we assume that the matter is cold. For completeness, we assume that the opacity and the EOS can be expressed in the form of power laws following the notations in~\cite{HR}, to obtain a self-similar solution:
\begin{subequations} \label{pwrlaws}
	\begin{equation}
	\frac{1}{\kappa_R}=gT^\alpha\rho^{-\lambda}
	\end{equation}
	\begin{equation} \label{pwrlaw_energy}
	e=fT^\beta\rho^{-\mu}
	\end{equation}
	\begin{equation}
	P(\rho,T)=r\rho e(\rho,T)\equiv(\gamma-1)\rho e(\rho,T)
	\label{def_r}
	\end{equation}    
\end{subequations}
where the temperature $T$ is measured in [heV] (1heV=100eV), and the density $\rho$ is measured in $[\mathrm{g/cm^3}]$. Eq.~\ref{def_r} implies that the EOS is assumed to be well described by an ideal gas, with an adiabatic index $\gamma\equiv r+1$. Furthermore, we assume that the boundary temperature is given as a power-law in time $T_s(t)=T_0\left(\frac{t}{t_s}\right)^{\tau}$,
where $t_s$ is a typical time scale ($1 \mathrm{nsec}$ in this work). 

When the heat wave propagates faster than the speed of sound, hydrodynamics is negligible, and we left with a simplified form of Eq.~\ref{energy_s} (without the $dV/dt$ term) along with Eqs.~\ref{pwrlaws}(a-b). This heat transfer is then characterized as of {\em{supersonic}} flow. In this case, the exact expression for the heat front position can be obtained from the self-similar solution (from~\cite{Garnier,shussman}):
\begin{equation}
\label{x_f_super} x_f(t)=\xi_f\sqrt{\frac{16}{3\left(4+\alpha\right)}\frac{g\sigma{T_0}^{4+\alpha-\beta}}{f\rho_0^{2+\lambda-\mu}}t_s}\left(\frac{t}{t_s}\right)^{\frac{1+\left(4+\alpha-\beta\right)\tau}{2}}
\end{equation}    
where $\xi_f$ is a dimensionless self-similarity parameter of order unity (see later in Fig.~\ref{fig:constants_xsi_z}(a)). For constant temperature BC ($\tau=0$), it yields the well-known $x_f\propto\sqrt{t}$ diffusion relation. $x_f(t)$ can be well approximated up to 1-3\% by the solution in~\cite{HR}, based on perturbation theory for a general profile of $T_s(t)$. For power-law thermal boundary condition it yields $\xi_f^2\approx\frac{2+\varepsilon}{(1-\varepsilon)[1+\tau(4+\alpha)]}$, which is accurate up to 1-5\%, where $\varepsilon\equiv\beta/(4+\alpha)$ (see also Fig.~\ref{fig:constants_xsi_z}(a)). Derivation of Eq.~\ref{x_f_super} in time yields the heat front velocity that scales like $\dot{x}_f(t)\propto{t^{\frac{\left(4+\alpha-\beta\right)\tau-1}{2}}}$. 

\begin{figure*}[htbp!]
\centering 
{
\includegraphics[width=12cm]{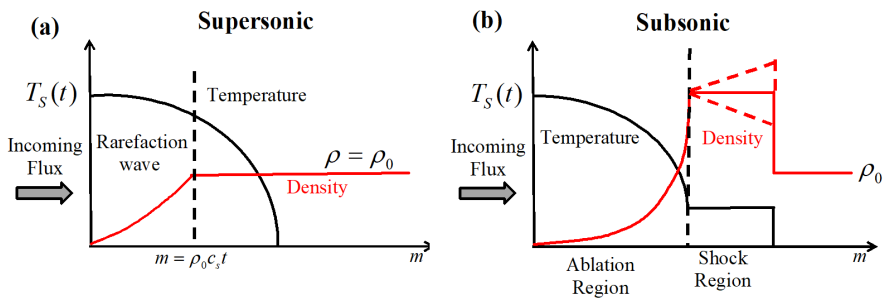}
}
\caption{Schematic temperature and density profiles as a function of the Lagrangian spatial coordinate for (a) a supersonic heat wave ($M>1$), including a rarefaction wave-zone and (b) a subsonic heat wave ($M<1$). The profiles in the shock wave region may have different spatial shapes (dashed lines).}  
\label{fig:super_sub_scheme}
\end{figure*}

For evaluating whether the radiative heat waves can be classified as supersonic or subsonic waves, it is convenient to use the definition of the Mach number ($\M$) as the ratio between the heat wave velocity and the speed of sound in the warm matter. Since the temperature profile behind the heat front is approximately isothermal, using Eqs.~\ref{pwrlaws}(b-c) the isothermal speed of sound is
$c_s(t)\equiv\sqrt{\left.\partial{P}/\partial{\rho}\right|_{T}}=\sqrt{r(1-\mu)fT^\beta\rho^{-\mu}}\propto{t}^{\frac{\beta\tau}{2}}$. Therefore, the Mach number is:
\begin{align}
\label{Mach_super} \M(t)=\frac{\dot{x}_f(t)}{c_s(t)}=\frac{1+\left(4+\alpha-\beta\right)\tau}{2}\xi_f\sqrt{\frac{16}{3(4+\alpha)(1-\mu)r}\frac{g\sigma{T_0}^{4+\alpha-2\beta}}{f^2\rho_0^{2+\lambda-2\mu}t_s}}\left(\frac{t}{t_s}\right)^{\frac{\left(4+\alpha-2\beta\right)\tau-1}{2}}
\end{align}  
These notations enable the estimation of the radiative hydrodynamic flows: supersonic, subsonic and intermediate/transition regimes. When $\M>1$ the heat wave is considered to be supersonic, and besides a rarefaction wave (which propagates at the speed of sound) near the front, the hydrodynamic motion is considered to be negligible (see Fig.~\ref{fig:super_sub_scheme}(a)). For $\M<1$ the heat wave is considered to be subsonic, when a shock wave propagating in the matter and ablation region is created behind the shock (see Fig.~\ref{fig:super_sub_scheme}(b)). In this work we focus on the transition region between supersonic heat flow to subsonic heat flow, which is characterized by $\M\approx1$.    

The extreme supersonic and subsonic regions were both examined and analyzed in many previous works (see~\cite{Marshak,zeldovich,ps,ps2,ger3,rosenScale3,rosenScale2,HR,Garnier,shussman,menahem}). The purely radiative transfer (supersonic) problem was introduced by~\citet{Marshak} and by Zel'dovich et al.~\cite{zeldovich}. An exact fully analytic solution for the temperature profile can be found only for two specific $\tau$'s: The Henyey solution for the case that $x_f$ increases linearly in time, and the Zel’dovich solution for energy impulse. However, an approximate (Henyey-like) solution may introduced for estimating the temperature profile in high-accuracy, knowing $x_f$ from Eq.~\ref{x_f_super} (or for general temperature profile using~\cite{HR}). For a wider discussion, see Appendix A. For the subsonic case, a solution for the ablation region hydrodynamics-involved was introduced by Pakula and Sigel, assuming infinite density at the heat front~\cite{ps,ps2,ger3}. A full self similar solution that includes both the radiation transport and the hydrodynamic motion cannot be achieved since the problem contains two characteristic velocity scales, the heat wave velocity and the speed of sound which are not self-similar altogether. An approximate patching-based approach was offered by~\citet{shussman}, and enabled analytic modeling of subsonic shock wave velocity experiments~\cite{alum}. Another approach was offered by Hammer and Rosen~\cite{HR} via perturbation theory, when a series of supersonic Marshak waves experiments is analyzed via their findings~\cite{avner1,avner2}. Recently,~\citet{menahem} has expanded the Zel'dovich supersonic solution for a non-homogeneous medium with a power law spatial density profile.

The transition region, or the intermediate regime, involves a non-negligible hydrodynamics motion and thus does not have an exact full-similar solution, due to the two characteristic velocity scales. Hence, Most of the solutions that were found in those previous works are not valid in the transition region. An exception of this aspect is presented in a milestone work by~\citet{Garnier}, which provides a unique self-similar solution for the whole physical problem, for a private case (a specific power-law behavior of the temperature boundary condition, "the critical case"), when the heat wave velocity and the speed of sound have the same temporal dependence. Hence, in this case, the Mach number is time independent, and the solution is valid for all values of $\M$. Another theoretical study of the intermediate regime, offers corrected analytic approximations for the total energy absorbed in the matter, which is important for the evaluation of the expected wall energy losses inside hohlraums in HEDP experiments~\cite{HR_PRE}. The study shows the existence of an optimal density for a minimizing the wall energy losses. This hypothesis was also examined experimentally~\cite{exp_PRL} (also qualitatively in~\cite{shay_drive_paper}), showing higher radiation temperatures in the low-density hohlraums rather than the solid-density ones. However, the theoretical analysis in~\cite{HR_PRE} is limited for constant temperature boundary condition.

The intermediate region was studied also experimentally, trying to characterize its unique hydrodynamic properties. The first major attempt is described in the works of Hoarty et al., using $50\mathrm{mg/cm^3}$ $\mathrm{C_9H_3O_2Cl_5}$ foams~\cite{hoarty1,hoarty2,hoarty3}. Another study used a $160\mathrm{mg/cm^3}$ $\mathrm{C_8H_8}$ foam~\cite{C8H8}, measuring both the heat and the shock fronts. Recently, a profound experimental study was published of the intermediate region using $50\mathrm{mg/cm^3}$ CHOBr foam~\cite{french_new}, showing the curvature of the shock front from the $\M\approx1$ to strong-shock conditions. The transition region in the last experiments~\cite{C8H8,french_new} occurs {\em{after}} the laser pulse ends, thus the temporal shapes of the temperature profile is complex, causing the analysis through self-similar solutions hard to obtain. Therefore, in this work we focus on the experiment that specifically intend to use the intermediate region {\em{during}} the laser pulse is on, the Young et al. experiment~\cite{exp_PRL}.

In this work, we theoretically analyze and characterize the transition region between supersonic and subsonic radiative heat flows. Since a general solution (even power-law) cannot be achievable as explained (besides numerical simulations), this work focuses on three goals: First, we introduce an {\em{approximate}} extended solution on for a general power-law temperature-dependency, based on~\citet{Garnier} ``critical case" exact solution. Second, we introduce an approximate solution for the energies stored in the matter, again, for a general power-law temperature profile, as an extension to the constant temperature approximation of~\citet{HR_PRE}. Both these approximations are tested against a wide survey of numerical simulations. Third, we test the strength of the approximate solutions with comparison to the experimental results of~\citet{exp_PRL}. In Sec.~\ref{sec:physical model} we present the main physical models used in this work, when the main results of this work are shown in Sec.~\ref{sec:Main results}. Finally, in Sec.~\ref{sec:PRL_exp_results} we reproduce the experimental results from~\cite{exp_PRL}.

\section{The physical model and the main methodology}
 \label{sec:physical model}

The theoretical model is divided to two main parts: First, we generalize Garnier's et al. exact `critical case' solution~\cite{Garnier}, that was derived for a specific power-law dependency, to a general BC. Second, we generalize Rosen's et al.~\cite{HR_PRE} analytic expressions that were derived for constant temperature BC, to a a general power-law dependency BC. Both of these theoretical models are examined against numerical simulations.

Most of the analysis in this work is shown for gold parameters (opacity and EOS) since this is the material which most of both the theoretical and experimental studies use. Some of the analysis will use \TaO instead, for testing the theoretical work against the experiment of~\citet{exp_PRL}. The parameters of the opacity and EOS that are shown in Eqs.~\ref{pwrlaws} are specified in table~\ref{table:pwr_law_opac_eos_Au}. These parameters are obtained from fitting the opacity and the EOS to the power law form in Eqs.~\ref{pwrlaws}. For gold (Au), these values are taken from~\cite{HR} for the temperature range of $100-200\mathrm{eV}$, and for \TaO the values were fitted from a QEOS table~\cite{QEOS} for the EOS, and from a CRSTA table~\cite{Kurz2012, Kurz2013} for the opacity in the temperature range of $100-300\mathrm{eV}$ (see also~\cite{avner1}).
\begin{table}[!htb]
	\centering
	\caption{\bf Power law fits for the EOS and opacity of gold~\cite{HR} and \TaO (fitted values from a QEOS table~\cite{QEOS} for the EOS, and from a CRSTA table~\cite{Kurz2012, Kurz2013} for the opacity) in temperatures within $1-2\mathrm{heV}$ for gold and $1-3\mathrm{heV}$ for \TaO.}
	\label{table:pwr_law_opac_eos_Au} 
	\begin{tabular}{|c|c|c|} \hline 
		\multicolumn{1}{|c|}{{\bf Parameter}} &
		\multicolumn{1}{c|}{{\bf Gold}} &
		\multicolumn{1}{c|}{{\bf \TaO}} \\ \hline
		\ $f$ $\mathrm{[MJ/g]}$ & $3.4$ & $4.65$  \\ \hline
		\ $\beta$ & $1.6$ & $1.37$  \\ \hline
		\ $\mu$ & $0.14$ & $0.12$  \\ \hline
		\ $g$ $\mathrm{[g/cm^2]}$ & $1/7200$  & $1/8200$  \\ \hline
		\ $\alpha$ & $1.5$ & $1.78$  \\ \hline
		\ $\lambda$ & $0.2$ & $0.24$  \\ \hline
		\ $r\equiv(\gamma-1)$ & $0.25$ & $0.29$  \\ \hline
	\end{tabular}
\end{table}

\subsection{Approximated derivation for a general $\tau$ -- extension of Garnier's et al. solution for the critical case $\tau_c$}
\label{garnier_extension}

As mentioned in the introduction, a unique exact full self-similar solution for both the ablation and the shock regions is introduced in~\cite{Garnier}, and it can be valid only for a specific $\tau=\tau_c$, also called ``the critical case". To achieve this full self-similar solution, the number of velocity scales in the problem has to be reduced from two to one. In other words, the heat front velocity and the speed of sound must have the same temporal dependency, which is equivalent to having a time-independent Mach number. From Eq.~\ref{Mach_super}, we have $\M(t)\propto{t^{\frac{\left(4+\alpha-2\beta\right)\tau-1}{2}}}$. Setting the exponent to zero and solving for $\tau$, yields (in~\cite{Garnier} the notation for $\tau_c$ is $k$):
\begin{equation} \label{tau_crit}
\tau\equiv\tau_c=\frac{1}{4+\alpha-2\beta},
\end{equation}
i.e., for this case, the Mach number is time independent, by definition, and stays constant in time. Hence, the Mach number in this case is a ``conserved quantity". Moreover, this critical case has another physical meaning, corresponds to a constant specific volume. 
Looking at the general form of the {\em{subsonic}} solution from~\cite{shussman} (also similar analysis in~\cite{ps,ps2}) for the specific volume, we can see that: $V(m,t)=1/\rho(m,t)\propto{t^{\frac{1-(4+\alpha-2\beta)\tau}{2+\lambda-2\mu}}}$, one can easily see that ``conservation of specific volume" yields $\tau=\frac{1}{4+\alpha-2\beta}\equiv\tau_c$. The meaning of conservation of the specific volume is that density profile stays invariant, and its shape does not change in time, for this specific $\tau$. This can be seen in a certain way as an equivalent requirement for having a self-similar solution for the whole problem, since in this case the initial density $\rho_0$ serves as the only density scale in the problem.

This unique self-similar solution for the critical case, is valid for both supersonic, intermediate and subsonic regimes. In appendix B we set a comparison between Garnier's solution in the critical cases for several values of $\M$, to the pure supersonic self-similar solution (for $\M>1$), and the deep subsonic (strong-shock) region, using Shussman \& Heizler's approximate solution (which was derived for general $\tau$)~\cite{shussman} for $\M<1$. The two solutions yield an excellent agreement as well as with numerical simulations, when in the supersonic case, the comparison is done of course for the temperature profile only, and in the deep subsonic case for all the hydrodynamical profiles. When the heat wave is weak-subsonic, and the shock is not a strong shock, the accuracy of Shussman \& Heizler's solution, decreases (for a wide discussion, see Appendix B). 

After introducing the physical meanings of the critical case, we derive an approximate solution for a general $\tau$. The derivation of the general $\tau$ case, starts with Eqs.~\ref{full_eqs_simple} (as in the critical case), with the following form (Ansatz):
\begin{subequations} \label{Ans_Garnier}
	\begin{equation} \label{ans_y}
	y\equiv{\frac{m}{m_f(t)}}, \quad \textrm{where:} \quad m_f(t)=m_{f0}\left(\frac{t}{t_s}\right)^{n}
	\end{equation}
	\begin{equation} \label{ans_T}
	T(t,y)=T_s(t)\tilde{T}(y)=T_0\left(\frac{t}{t_s}\right)^{\tau}\tilde{T}(y)
	\end{equation}
	\begin{equation} \label{ans_V}
	V(t,y)=V_s(t)\tilde{V}(y)=V_{s0}\left(\frac{t}{t_s}\right)^{\ell}\tilde{V}(y)
	\end{equation}
	\begin{equation}
	u(t,y)=u_s(t)\tilde{u}(y)=u_{s0}\left(\frac{t}{t_s}\right)^{j}\tilde{u}(y)
	\label{ans_u}
	\end{equation}    
\end{subequations} 
where $V_{s0}=1/\rho_0$. The boundary conditions are: $T(t,0)=T_s(t)$, $\rho(t,0)=0$, $T(t,1)=0$, $\rho(t,1)=\rho_0$, $u(t,1)=0$. Substituting the self-similar Ansatz into Eqs.~\ref{full_eqs_simple}, the exponents $n,\ell,j$ can be found from the requirement the equations will become dimensionless, in particular, without any temporal dependence on either sides of the equations. This Ansatz assumes the existence of a self-similar solution to the problem, and that the spatial profiles depends only on a single dimensionless coordinate $y$. However, this assumption is not valid for $\tau\ne\tau_c$, since there are more than one typical time-scale in the problem. Thus, the solution obtained followed by this Ansatz is not expected to be exact, except for the critical case. In the critical case, for $\tau=\tau_c$, the exponents are obtained to be: $n=1+\frac{\beta\tau_c}{2}$, $\ell=0$, $j=n-1$. $\ell=0$ implies constant specific volume/density profile in time, as mentioned earlier. the velocity exponent $j=\frac{\beta\tau_c}{2}$ reproduces the speed of sound temporal dependency.

For a general power-law boundary condition, the self-similar Ansatz yields the exponents $n,\ell,j$ as a function of $\tau$:
\begin{subequations} \label{exponents_Garnier_all_tau}
	\begin{equation} \label{exponents_n}
	n=\frac{2+2\lambda-3\mu+\left({(2-\mu)(4+\alpha)-(2-\lambda)\beta}\right)\tau}{2(2+\lambda-2\mu)}
	\end{equation}
	\begin{equation} \label{exponents_l}
	\ell=\frac{1-(4+\alpha-2\beta)\tau}{2+\lambda-2\mu}
	\end{equation}
	\begin{equation} \label{exponents_j}
	j=n+\ell-1=\frac{\mu+\left({(2+\lambda)\beta-\mu(4+\alpha)}\right)\tau}{2(2+\lambda-2\mu)}
	\end{equation}    
\end{subequations}   

The left unknown dimensional parameters are identified by the requirement that all the dimensional parameters must be reduced from the equations. The values are obtained to be: $m_{f0}=\left(rf{T_0}^\beta\rho_0^{2-\mu}\right)^{\nicefrac{1}{2}}t_s\xi_f$ and $u_{s0}=\frac{m_{f0}}{\rho_0t_s}$, where $\xi_f$ is the self-similarity parameter, which is an unknown. The dimensionless equations for the shape functions are:
\begin{subequations} \label{dimensionless_Garnier_all_tau}
	\begin{equation} \label{dimensionless_V_all_tau}
	\frac{d\ti{V}}{dy}=-\frac{\varepsilon\ti{\zeta}^{\varepsilon-1}\ti{V}^{\mu-1}\ti\theta+\xi_f^2(j\ti{u}-n\ell{y}\ti{V})}{n^2\xi_f^2y^2+(\mu-1)\ti{\zeta}^{\varepsilon}\ti{V}^{\mu-2}}
	\end{equation}
	\begin{equation} \label{dimensionless_u_all_tau}
	\frac{d\ti{u}}{dy}=\ell\ti{V}-ny\frac{d\ti{V}}{dy}
	\end{equation}
	\begin{equation} \label{dimensionless_zeta_all_tau}
	\frac{d\ti{\zeta}}{dy}=\ti{\theta}
	\end{equation}
	\begin{align}
	\label{dimensionless_psi_all_tau}        
	\frac{d\ti{\theta}}{dy}=\frac{\xi_f^2\ti{V}^{-\lambda}}{\chi_0} \bigg\{ & (\beta\tau+\ell(\mu+r))\ti{\zeta}^{\varepsilon}\ti{V}^{\mu}-\varepsilon{ny}\ti{\zeta}^{\varepsilon-1}\ti{V}^{\mu}\ti\theta- \nonumber \\
	& \left[\lambda\frac{\chi_0}{\xi_f^2}\ti{V}^{\lambda-1}\ti{\theta}+ny(\mu+r)\ti{\zeta}^{\varepsilon}\ti{V}^{\mu-1} \right] \frac{d\ti{V}}{dy}\bigg\}     
	\end{align}    
\end{subequations} 
where $\ti{\zeta}\equiv\ti{T}^{4+\alpha}$ and:
\begin{equation}
\label{chi0}
\chi_0\equiv\frac{16g\sigma{T_0}^{4+\alpha-2\beta}}{3(4+\alpha)rf^2\rho_0^{2+\lambda-2\mu}t_s}. 
\end{equation}
$\chi_0$ is a unique dimensionless parameter that depends on the material's properties, density, temperature, 
that comes out from reducing the original equations into their dimensionless form.
The self-similar solution is characterized entirely by the parameter $\chi_0$. Although is it not specified explicitly in~\cite{Garnier}, this parameter is very close related to the Mach number (Eq.~\ref{Mach_super}) as:
\begin{align}
\label{Mach_super_G} \M(t)=\frac{1+\left(4+\alpha-\beta\right)\tau}{2}\frac{\xi_f}{\sqrt{1-\mu}}\sqrt{\chi_0}\left(\frac{t}{t_s}\right)^{\frac{\left(4+\alpha-2\beta\right)\tau-1}{2}} 
\end{align}    
For $t=t_s$ the Mach number is simply $\M(t_s)=\frac{1+\left(4+\alpha-\beta\right)\tau}{2}\frac{\xi_f}{\sqrt{1-\mu}}\sqrt{\chi_0}$, which means that for this time, up to a constant factor of order unity, $\chi_0\propto{{\M^2(t_s)}}$. Thus, a supersonic heat wave implies $\chi_0\gtrapprox~1$, and a subsonic heat wave implies $\chi_0\lessapprox~1$. For $\chi_0\approx1$ the heat wave is close to the sonic point and it is in the transition region~\cite{Garnier}.

It is evident that the equations resemble of their original form shown in~\cite{Garnier}. The changes are: an additional term of $\ell\ti{V}$ in Eq.~\ref{dimensionless_u_all_tau}, an additional term of $n{\ell}y\xi_f^2\ti{V}$ in the numerator of Eq.~\ref{dimensionless_V_all_tau}, as well as a generalized form of the the term $\xi_f^2(n-1)\ti{u}$ in the original equation, which becomes $\xi_f^2j\ti{u}$ in Eq.~\ref{dimensionless_V_all_tau} (for the critical case $j=n-1$). The last change is in Eq.~\ref{dimensionless_psi_all_tau}, where the coefficient of $2(n-1)$ from the original equation is replaced by the expression $\beta\tau+\ell(\mu+r)$. All of those changes are terms which depend directly on $\ell$, thus they do not exist in the critical case since $\ell(\tau_c)=0$.

\begin{figure}[th]
	\centering
	\includegraphics[width=7.5cm]{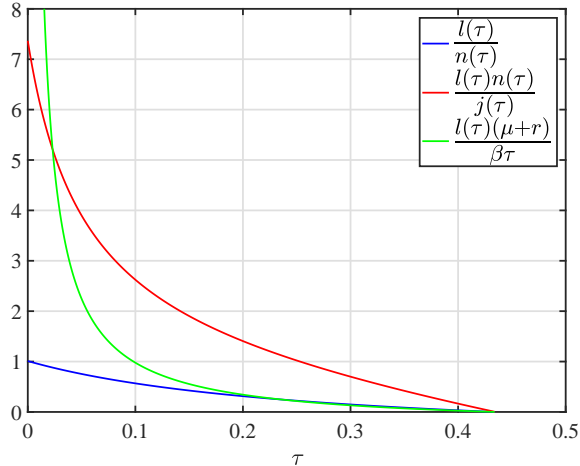}
	\caption{The relative difference between terms in the dimensionless equations, compared to their nominal value for the critical case as a function of $\tau$ for gold, the material parameters are taken from table~\ref{table:pwr_law_opac_eos_Au}}.
	\label{fig:ss_terms_ratios}
\end{figure}

Since we already know that for $\tau\ne\tau_c$, the solution of Eqs.~\ref{dimensionless_Garnier_all_tau} would not be exact, we wish to estimate, based on those changes, in the dimensionless equations, for what values of $\tau$ the solution of Eqs.~\ref{dimensionless_Garnier_all_tau} would still yield a good accuracy with the exact solution (of Eqs.~\ref{full_eqs_simple}). The relative difference between the additional new terms and their nominal value for $\tau=\tau_c$ is presented as a function of $\tau$ in Fig.~\ref{fig:ss_terms_ratios} for gold parameters (taken from table~\ref{table:pwr_law_opac_eos_Au}).
Since the equations are dimensionless and all those added terms serve as coefficients to other terms, of order of unity, we expect the added new terms to affect the solution significantly when the relative difference is of the same order of magnitude. From Fig.~\ref{fig:ss_terms_ratios} it can be seen, that the term with the highest relative ratio for a given $\tau$ is the $n{\ell}y\xi_f^2\ti{V}$ term, in Eq.~\ref{dimensionless_V_all_tau}. For $\tau=0.25$ the relative ratio for this term is close to $1$, meaning the terms are approximately equally important. Thus, we assume that for $\tau\le{0.25}$ the obtained solution of Eqs.~\ref{dimensionless_Garnier_all_tau} might yield large errors. On the contrary, for $\tau>{0.25}$ this solution is expected to be quite close to the exact solution.

The boundary conditions of the original (dimensional) equations imply dimensionless boundary conditions on the shape functions at $y=0$ and $y=1$: $\ti{\zeta}(0)=1$, $\lim_{y\to 0} \ti{V}=\infty$, $\ti{\zeta}(1)=0$, $\ti{V}(1)=1$, $\ti{u}(1)=0$. These boundary conditions complete the set of dimensionless equations, as we have four first order ODE for four unknown shape functions and another unknown self-similarity parameter ($\xi_f$), along with five boundary conditions. Naively, this set of equations is solved numerically, via a numeric solver integrating the equations from $y=1$ to $y=0$ and using a shooting method to verify the boundary conditions at $y=0$ are met (same as in~\cite{Garnier}). However, solving the set of equations in their current form, by starting the integration from $y=1$ backward to $y=0$ turns out to be impossible, as mentioned in~\cite{Garnier}. Mathematically that is because direct integration of Eq.~\ref{dimensionless_V_all_tau} leads to a singularity, as the denominator vanishes for some $y_0<1$. From a physical point of view, we have to consider the profiles are not continuous, due to the shock wave ahead of the heat wave for the subsonic case, or the isothermal shock behind the heat front for the supersonic case. To deal with this issue, some new unknown parameter are introduced - $y_c$ for the supersonic/transition region,
that denotes the dimensionless position of the isothermal shock. For the (deep) subsonic region, for low enough $\chi_0$, the shock is a strong classical one, $y_a$ is introduced instead, when $y=1$ denotes the dimensionless position of the shock wave, and $y_a$ denotes the dimensionless position of the ablation front. We follow the numeric method of solution offered by Garnier et al.~\cite{Garnier}, with the adjustments that need to be done for general $\tau$. 

First, the numeric solution algorithm, for the supersonic/transition region is presented (for large/intermediate values of $\chi_0$). Since there is a shock wave at $y_c$, we need to specify the jump conditions at the shock front. Those can be obtained via direct integration of Eqs.~\ref{dimensionless_Garnier_all_tau} for a small interval near $y_c$, i.e., from $y_c^{-}<y_c$ to $y_c^{+}>y_c$. Integration over full derivatives gives the jump of the integrand, while integration of expression which does not contain derivatives vanishes in the limit of $y_c^{+}-y_c^{-}\to 0$. The obtained relations for the jump conditions at $y_c$ are:
\begin{subequations} \label{Garnier_tau_c_yc_BC}
	\begin{equation} \label{Garnier_tau_c_yc_BC_zeta}
	\ti{\zeta}(y_c^{-})=\ti{\zeta}(y_c^{+})
	\end{equation}
	\begin{equation} \label{Garnier_tau_c_yc_BC_u}
	\ti{u}(y_c^{-})=\ti{u}(y_c^{+})+ny_c[\ti{V}(y_c^{+})-\ti{V}(y_c^{-})]
	\end{equation}
	\begin{align} \label{Garnier_tau_c_yc_BC_theta}
	\ti{\theta}(y_c^{-})=&\ti{\theta}(y_c^{+})\frac{\ti{V}^{\lambda}(y_c^{+})}{\ti{V}^{\lambda}(y_c^{-})}-\frac{\xi_f^2}{\chi_0\ti{V}^{\lambda}(y_c^{-})} \bigg\{  ny_cr\frac{\xi_f^2}{2}[\ti{u}^2(y_c^{-})-\ti{u}^2(y_c^{+})]+ \nonumber \\ 
	& +ny_c\ti{\zeta}^{\varepsilon}(y_c)[\ti{V}^{\mu}(y_c^{-})-\ti{V}^{\mu}(y_c^{+})] \nonumber \\ 
	& -r\ti{\zeta}^{\varepsilon}(y_c)[\ti{V}^{\mu-1}\ti{u}(y_c^{-})-\ti{V}^{\mu-1}\ti{u}(y_c^{+})] \bigg\}    
	\end{align}        
	\begin{align} \label{Garnier_tau_c_yc_BC_V}    
	&\textrm{as for the specific volume, the jump is expressed implicitly:} \nonumber \\ 
	&\psi(\ti{V}(y_c^{-}))=\psi(\ti{V}(y_c^{+})) \textrm{,} \\ 
	&\textrm{where the function} \quad \psi(V)  \quad \textrm{is defined by:} \nonumber \\
	&\psi(V)=\ti{\zeta}^{\varepsilon}(y_c)V^{\mu-1}+\xi_f^2n^2{y_c}^2V \nonumber
	\end{align}                
\end{subequations} 
Eq.~\ref{Garnier_tau_c_yc_BC_zeta} thus shows that the shock is indeed isothermal, since $\ti{\zeta}=\ti{T}^{4+\alpha}$. Also, a unique distinct jump from $\ti{V}(y_c^{+})$ to $\ti{V}(y_c^{-})$ is guaranteed, since the function $\psi(V)$ always have a minimum point between the two points $\ti{V}(y_c^{+})$ and $\ti{V}(y_c^{-})$. The jump conditions remain unchanged comparing to the critical case, since none of the new terms in the dimensionless equations contains derivatives of the shape functions, hence, they all vanish after the integration, taking the limit of $y_c^{+}-y_c^{-}\to 0$. The solution is found in several steps. First, the system is integrated from $y=1$ to $y=y_c^{+}$ for some initial guess for the parameter. Then the jump conditions are applied at the guessed $y_c$. In the last step, the integration of the Eqs. continues from $y=y_c^{-}$ to $y=0$, while the values of $\xi_f,y_c$ are found via a double shooting method in iterations, until at the end of the process, the boundary conditions at $y=0$ are met.

Finally, the numeric solution algorithm, for the subsonic region is presented (for small values of $\chi_0$). We solve Eqs.~\ref{dimensionless_Garnier_all_tau} from $y_a^{-}$ to $y=0$, but from $y=1$ to $y_a^{+}$ a simplified version is solved, only for the hydrodynamics, neglecting the radiative heat conduction, as in~\cite{Garnier} (also similarly to Shussman and Heizler's rationale~\cite{shussman}). The simplified dimensionless ($\tau$-dependent) hydrodynamics equations for the $y_a^{+}\leqslant y \leqslant 1$ region are:
\begin{subequations} \label{dimensionless_Garnier_all_taus_shock}
	\begin{equation} \label{dimensionless_V_all_taus_shock}
	\frac{d\ti{V}}{dy}=-\frac{(\beta\tau+\ell(\mu+r))\ti{\zeta}^{\varepsilon}\ti{V}^{\mu-1}+n\xi_f^2y(j\ti{u}-n{\ell}y\ti{V})}{ny[n^2\xi_f^2y^2-(1+r)\ti{\zeta}^{\varepsilon}\ti{V}^{\mu-2}]}
	\end{equation}
	\begin{equation} \label{dimensionless_u_all_taus_shock}
	\frac{d\ti{u}}{dy}=\ell\ti{V}-ny\frac{d\ti{V}}{dy}
	\end{equation}
	\begin{equation} \label{dimensionless_zeta_all_taus_shock}
	\frac{d\ti{\zeta}}{dy}=\frac{\beta\tau+\ell(\mu+r)}{n\varepsilon{y}}\ti{\zeta}-\frac{r+\mu}{\varepsilon}\ti{\zeta}\ti{V}^{-1}\frac{d\ti{V}}{dy}
	\end{equation}
\end{subequations}   
Eqs.~\ref{dimensionless_Garnier_all_taus_shock} contain few changes compared to the original equations in~\cite{Garnier} for $\tau=\tau_c$. In Eq.~\ref{dimensionless_u_all_taus_shock} there is an added term of $\ell\ti{V}$, in the numerator of Eq.~\ref{dimensionless_V_all_taus_shock} the coefficient $2(n-1)$ for the critical case is replaced by $\beta\tau+\ell(\mu+r)$, and there is also an added term of $n^2\ell\xi_f^2y^2\ti{V}$. In Eq.~\ref{dimensionless_zeta_all_taus_shock} the coefficient was replaced from $4n-3$ in the critical case to $\frac{\beta\tau+\ell(\mu+r)}{\varepsilon}$.

The boundary conditions for a general $\tau$ remain the same as for $\tau_c$. Since the dimensionless position of the shock front is set to $y=1$, the jump conditions for a classical strong shock should be applied there. By Rankine-Hugoniot relations, relying on the values ahead of the shock at $y=1^{+}$ from the original boundary conditions at $y=1$, it holds that: 
\begin{subequations} \label{Garnier_tau_c_shock_BC}
	\begin{equation} \label{Garnier_tau_c_shock_BC_V}
	\ti{V}(1^{-})=\frac{r}{r+2}    
	\end{equation}    
	\begin{equation} \label{Garnier_tau_c_shock_BC_u}
	\ti{u}(1^{-})=\frac{2n}{r+2}
	\end{equation}
	\begin{equation} \label{Garnier_tau_c_shock_BC_zeta}
	\ti{\zeta}(1^{-})=\left(\frac{2n^2\xi_f^2r^{1-\mu}}{(r+2)^{2-\mu}}\right)^{1/\varepsilon}
	\end{equation}    
\end{subequations}     
The solution for that case found in a quite similar way to the previous one. First, the jump conditions are applied at $y=1^{-}$. Then, the simplified system of Eqs.~\ref{dimensionless_Garnier_all_taus_shock} is solved by integration from $y=1^{-}$ to $y=y_a^{+}$, for an initial guess for $y_a$. Finally, the full system of Eqs.~\ref{dimensionless_Garnier_all_tau} is solved by integration from $y=y_a^{-}$ to $y=0$ (the solution is continuous at $y_a$). The process continues in iterations again, via a double shooting method, until the correct values of $\xi_f,y_a$ are found, so that the boundary conditions at $y=0$ are met.

A demonstration of hydrodynamical profiles using the extended Garnier's solution for the different cases (using several values of $\chi_0$) is presented in the next section (see for example Figs.~\ref{fig:profiles_Au_all_taus_T0=100_t=1} and~\ref{fig:profiles_Au_sesame_high_density_T0=100_t=1} or in Appendix B, Fig.~\ref{fig:profiles_Au_tau_c_T0=100_t=1}). A detailed comparison of this solution results with numerical simulations results is also performed. Such a comparison will enable an examination of the quality of the derived approximation.

\subsection{Approximated solutions for the absorbed energies for a general $\tau$ -- extension of Rosen's et al. solution for $\tau=0$}
\label{approximated}

In this subsection we present an extension of the approximated study of~\cite{HR_PRE} that was derived for a constant temperature BC ($\tau=0$), for a general $\tau$. We derive analytic approximations for the absorbed total energy (per area unit) within the matter which examines the dependency of the total energy on the initial density of the matter $\rho_0$, for a general BC. The derivation here is done for general material parameters, so one can use this approach to model other problems as well. We then present two quantitative physical examples, gold, which is the material which most of the similar studies use (usually for hohlraum experiments)~\cite{HR_PRE,HR}, for example, and for \TaO, which is the material used in Young's experiment~\cite{exp_PRL}. We present here the full approximation for the energy that relies on the results from the self-similar solutions~\cite{Garnier,shussman}, for both the supersonic and the subsonic regimes:
\begin{subequations}
\begin{equation} \label{E_A_sup_pure}
E_{\mathrm{Super,pure}}/A=E_{0,\mathrm{Super}}{T_0}^{p_{T,\mathrm{Super}}}{\rho_0}^{p_{\rho,\mathrm{Super}}}{t}^{p_{t,\mathrm{Super}}}  
\end{equation}
\begin{equation} \label{E_A_sub_pure}
E_{\mathrm{Sub,pure}}/A=E_{0,\mathrm{sub}}{T_0}^{p_{T,\mathrm{Sub}}}{t}^{p_{t,\mathrm{Sub}}}  
\end{equation}
\end{subequations}
where $E_{0,\mathrm{Super}}$/$E_{0,\mathrm{Sub}}$ is a constant that is determined from the dimensionless equation, and it is a $\tau$-dependent, $p_{T,\mathrm{Super}}$, $p_{\rho,\mathrm{Super}}$ and $p_{t,\mathrm{Super}}$ are the powers for the temperature, density and time powers for the supersonic case, respectively, while $p_{T,\mathrm{Sub}}$ and $p_{t,\mathrm{Sub}}$ are the powers for the  subsonic case, and they are determined from a dimensional-analysis (there is no density-dependency in the subsonic case).

In the supersonic case, considering that the self-similar solution assumes that the hydrodynamic is negligible, the total energy per area unit is given by: $E_{\mathrm{super, pure}}(t)/A=\int_{0}^{m_f(t)}{e(m,t)dm}=\rho_0\int_{0}^{x_f(t)}{edx}$. Substitution of the EOS form Eq.~\ref{pwrlaws} and the self-similar solution form for $T(x,t)$ (Eq.~\ref{ans_T}) and $x_f(t)$ (Eq.~\ref{x_f_super}), we obtain: 
\begin{equation}
\label{E_super_1}
E_{\mathrm{Super,pure}}(t)/A=\sqrt{\frac{16\sigma{fg}t_s}{3(4+\alpha)}}{\rho_0}^{-\frac{\mu+\lambda}{2}}{T_0}^{\frac{4+\alpha+\beta}{2}}{\left(\frac{t}{t_s}\right)}^{\frac{1+(4+\alpha+\beta)\tau}{2}}{\xi_{f,\mathrm{Super}}\ti{z}_{\mathrm{Super}}}
\end{equation}
where: $\ti{z}_{\mathrm{super}}=\int_{0}^{1}{{\ti{T}(y)}^{\beta}dy}$ is the dimensionless energy, and $y=x/x_f(t)$. The coefficients $\xi_{f,\mathrm{Super}}$ and $\ti{z}_{\mathrm{Super}}$ are of order unity and are determined from the dimensionless ODE solution, and are functions of $\tau$ in the general case. The solution for $\ti{z}_{\mathrm{Super}}$ can be also well approximated by the solution in~\cite{HR}, based on perturbation theory, for a general profile of $T_s(t)$, $\ti{z}_{\mathrm{Super}}\approx1-\varepsilon=\frac{4+\alpha-\beta}{4+\alpha}$. This approximated result {\em{is not}} a function of $\tau$, and its accuracy is up to 1-5\%.

\begin{figure}[htp]
	\centering
	(a)
	\includegraphics[width=7.5cm,clip=true]{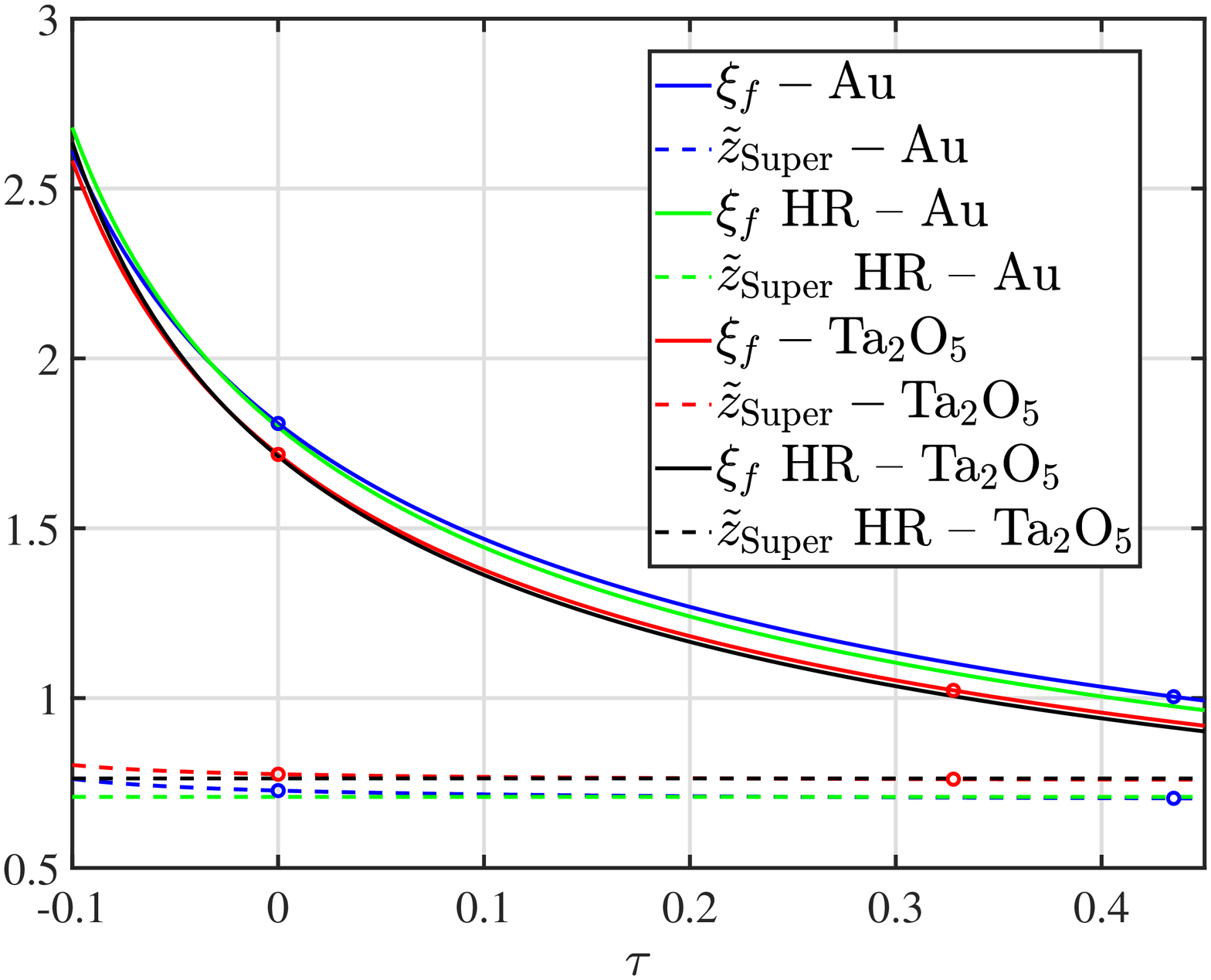}
	(b)
	\includegraphics[width=7.5cm,clip=true]{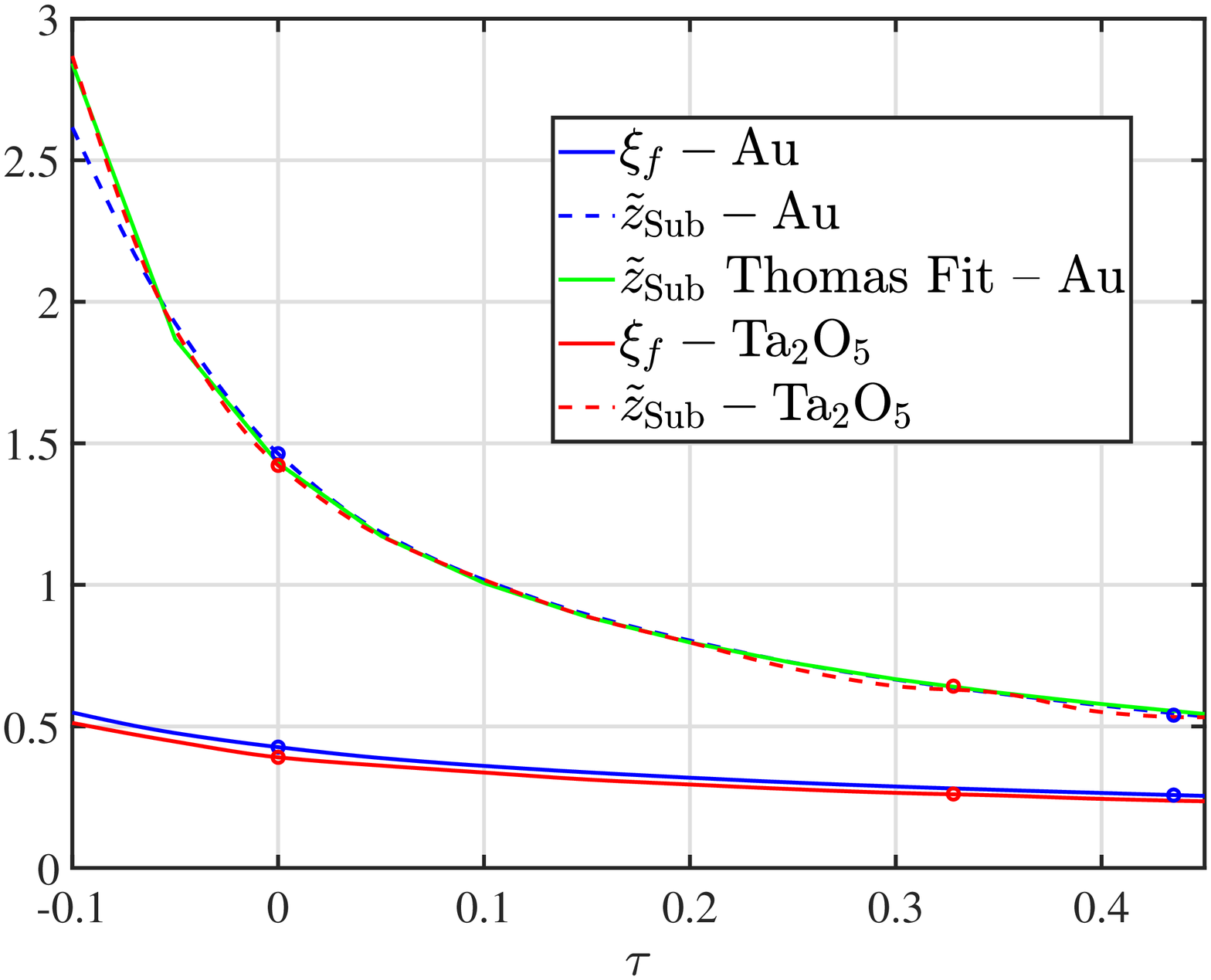}
	\caption{The dimensionless self similar heat front parameter $\xi_f$ (solid curves) and the dimensionless energy $\tilde{z}$ (dashed carves) for the supersonic (a) and for the subsonic case (b). The exact self-similar values are in blue for Au and in red for \TaO. The green (Au) and black (\TaO) curves in (a) show the values obtained from the perturbation theory solution in~\cite{HR}. The green curve in (b) shows the fitted expression for the dimensionless energy from~\citet{Thomas2008}.}
	\label{fig:constants_xsi_z}
\end{figure}

In Fig.~\ref{fig:constants_xsi_z}(a) these constants are plotted for gold parameters (blue curves) and for \TaO (red). In addition, we plot~\citet{HR} perturbation theory solution for both gold (green) and \TaO (black). The matching between the approximated and the exact $\xi_{f,\mathrm{Super}}$ and $\ti{z}_{\mathrm{Super}}$ is very good, up to 5\%. Specifically, the exact $\ti{z}_{\mathrm{Super}}$ varies very weakly with $\tau$, as the Hammer's \& Rosen's solution predicts.

In the subsonic case, the total energy per area unit is given by: $E_{\mathrm{sub, pure}}(t)/A=\int_{0}^{m_f(t)}{\left(e(m,t)+\frac{1}{2}u^2(m,t)\right)dm}$. Using the results for the self-similar solutions in~\cite{Garnier,shussman} it can be written as: 
\begin{align}
\label{E_sub_1}
E_{\mathrm{Sub,pure}}(t)/A=& \Bigg(\left({\frac{16\sigma{g}}{3(4+\alpha)\xi_{f,\mathrm{Sub}}}}\right)^{2-3\mu}{r}^{\mu+\lambda}{f}^{2+3\lambda}{T_0}^{2(4+\alpha+\beta)+3\lambda\beta-3\mu(4+\alpha)}\nonumber \\ &{t}^{2+2\lambda-\mu+[2(4+\alpha+\beta)+3\lambda\beta-3\mu(4+\alpha)]\tau}\Bigg)^{\frac{1}{4+2\lambda-4\mu}}\ti{z}_{\mathrm{Sub}} 
\end{align}
where: $\ti{z}_{\mathrm{Sub}}=\int_{0}^{1}{\left({\ti{T}(y)}^{\beta}{\ti{V}(y)}^{\mu}+\frac{r}{2}{\ti{u}(y)}^{2}\right)dy}$ is the dimensionless energy. There are not simple analytic approximate expressions for the subsonic case, however, for gold, a simple analytic fit can be found for $E_{\mathrm{Sub,pure}}(t)/A$ (i.e., for $\ti{z}_{\mathrm{Sub}}$ via Eq.~\ref{E_sub_1}) by~\citet{Thomas2008}, $E_{\mathrm{Thomas}}=\frac{3.4(1+4\tau)}{3.35\tau+0.59}\mathrm{[kJ/cm^2]}$. In Fig.~\ref{fig:constants_xsi_z}(b) $\xi_{f,\mathrm{Sub}}$ and $\ti{z}_{\mathrm{Sub}}$ are plotted for gold parameters (blue curves) and for \TaO (red). Thomas fit is plotted (for gold, in green), where the matching is very good, especially for $\tau>0$. $\ti{z}_{\mathrm{Sub}}$ of gold and \TaO are very similar quantitatively.

Deriving the fixed energies of~\citet{HR_PRE} for a general $\tau$ always starts with the values of $\ti{z}_{\mathrm{Super/Sub}}$ that is presented Fig.~\ref{fig:constants_xsi_z}. For example, for $\tau=0$, the {\em{supersonic}} stored energies for gold and \TaO are:
\begin{subequations}
\begin{equation} \label{E_super_pure_gold}
{E_{\mathrm{Au,Super,pure}}}/A=2.86{T_0}^{3.55}{\rho_0}^{-0.17}{t}^{0.5}\quad\mathrm{[kJ/cm^2]}
\end{equation} 
\begin{equation} \label{E_super_pure_ta}
{E_{\mathrm{Ta_2O_5,Super,pure}}}/A=3.09{T_0}^{3.575}{\rho_0}^{-0.175}t^{0.5}\quad\mathrm{[kJ/cm^2]} 
\end{equation} 
\end{subequations}
The values for Au in Eq.~\ref{E_super_pure_gold} is very similar to (reproduces) the values in Eq. (3) in~\cite{HR_PRE}. The pure {\em{subsonic}} stored energies for gold and \TaO are for $\tau=0$,:
\begin{subequations}
\begin{equation} \label{E_sub_pure_gold}
E_{\mathrm{Au,Sub,pure}}/A=5.92{T_0}^{3.35}t^{0.59}\quad\mathrm{[kJ/cm^2]}
\end{equation} 
\begin{equation} \label{E_sub_pure_ta}
E_{\mathrm{Ta_2O_5,Sub,pure}}/A=7.21{T_0}^{3.31}t^{0.59}\quad\mathrm{[kJ/cm^2]}
\end{equation} 
\end{subequations}
Again, the values for Au in Eq.~\ref{E_sub_pure_gold} is very similar to (reproduces) the values in Eq. (5) in~\cite{HR_PRE}. For any other value of $\tau$ one can use Fig.~\ref{fig:constants_xsi_z} (or solving explicitly the ODE solution/using one of the offered approximations).

Next, we take care of the correction term for the supersonic regime, due to the hydrodynamic rarefaction wave.
For $\M\gtrsim1$ (near the sonic point) the hydrodynamic motion cannot be neglected, due to the rarefaction wave that propagates in the matter, behind the heat front. Following the rationale of~\cite{HR_PRE}, to calculate the total energy due to the rarefaction wave, the hydrodynamic profiles for the density, temperature, and velocity must be calculated via the hydrodynamic equations of motion (Eqs.~\ref{full_eqs_simple}). Simplifying the problem, one assumes the rarefaction wave is {\em{spatially}} isothermal, since the temperature profile behind the heat front in the supersonic case is quite flat (see also Appendix A). The density and velocity profiles are found from the simplified equations for mass and momentum conservation (as in~\cite{Garnier}):
\begin{subequations}
	\label{hydro_eqs}
	\begin{equation}
	\label{hydro_mass}
	\p{\rho}{t}+\p{}{x}(\rho{u})=0
	\end{equation}
	\begin{equation}
	\label{hydro_momentum}
	\p{u}{t}+u\p{u}{x}+\frac{1}{\rho}\p{P}{x}=0
	\end{equation}                            
\end{subequations}  
The initial conditions are: $\rho(x,t=0)=\rho_0, u(x,t=0)=0$, and boundary conditions at the front of the rarefaction wave, $x_r$, are: $\rho(x_r,t)=\rho_0, u(x_r,t)=0$. Eqs.~\ref{hydro_eqs} can be solved with a self-similar solution~\cite{Garnier}, assuming $\tau=0$, since it is the only value of $\tau$ for which a simple analytic solution can be found. Hence, we first consider the correction for a constant boundary temperature ($\tau=0$), and based on this result we will generalize it later for all values of $\tau$. For $\tau=0$ (and general $\mu$) the obtained solution is:
\begin{subequations}
\label{exact_rare}
\begin{equation}
{\rho}(x,t)=\rho_0\left(1+\frac{\mu}{2-\mu}\left(1-\frac{x}{c_0t}\right)\right)^{-\frac{2}{\mu}}
\end{equation}
\begin{equation}
{u}(x,t)=\frac{2c_0}{2-\mu}\left(\frac{x}{c_0t}-1\right)
\end{equation}
\end{subequations}
where: $c_0=\sqrt{(1-\mu)rf{T_0}^\beta{\rho_0}^{-\mu}}$ is the isothermal speed of sound for $\tau=0$). For $\mu=0$, the solution approaches the classical exponential solution that was used in~\citet{HR_PRE}:
\begin{subequations}
\label{approx_rare}
\begin{equation}
\rho(x,t)=\rho_0\exp(-1+{x}/{c_0t})
\end{equation}
\begin{equation}
{u}(x,t)=c_0({x}/{c_0t}-1)
\end{equation}
\end{subequations}
As opposed to~\cite{HR_PRE}, this work uses the general solution of Eq.~\ref{exact_rare}.

The derivation from Eqs.~\ref{delta_kinetic} to~\ref{delta_total} is similar to Secs. II(G-H) in~\citet{Garnier}. Since the hydrodynamic profiles end at the front of the rarefaction wave $x_r(t)$, for any given time $t$, we calculate the additional energy only on this segment. Thus, the additional kinetic energy is:
\begin{equation}
\Delta{E_k}/A=\int_{-\infty}^{c_0t}{\frac{1}{2}\rho(x,t)u^2(x,t)dx}=\frac{2}{(1-\mu)(2-3\mu)}\rho_0c_0^3t
\label{delta_kinetic}
\end{equation}
Similarly, the internal energy as a result of the rarefaction wave is given by:
\begin{equation}
\Delta{E_{i,r}}/A=\int_{-\infty}^{c_0t}{\rho(x,t)e(x,t)dx}=\frac{2-\mu}{r(1-\mu)(2-3\mu)}\rho_0c_0^3t
\end{equation}
Nevertheless, the original expression for the energy in the supersonic case has already counted for the internal energy at the rarefaction wave spatial region. To avoid ``double counting" we subtract from $\Delta{E_{i,r}}$ the amount of energy for $x\leqslant{x_r(t)}$ while considering there is no hydrodynamic motion. This Energy difference is given by:
\begin{equation}
\Delta{E_{i,r'}}/A=\int_{-\infty}^{c_0t}{\rho(x,t)e(x,t)dx}=\frac{1}{r(1-\mu)}\rho_0c_0^3t,
\end{equation}
since $\rho(x,t)=\rho_0$. 
Thus, the total additional energy is:
\begin{equation}
\Delta{E_{\mathrm{Super,tot}}}/A=\left(\Delta{E_{k}}/A+\Delta{E_{i,r}}/A\right)-\Delta{E_{i,r'}}/A=\frac{2(\mu+r)}{r(1-\mu)(2-3\mu)}\rho_0c_0^3t,
\label{delta_total}
\end{equation}
and the total energy stored in the matter in the supersonic regime is: 
\begin{equation}
\label{final_super_cor}
E_{\mathrm{Super,tot}}/A=E_{\mathrm{Super,pure}}/A+\Delta{E_{\mathrm{Super,tot}}}/A,
\end{equation}
using $E_{\mathrm{Super,pure}}/A$ from Eq.~\ref{E_super_1}.

So far the derivation was similar to~\cite{HR_PRE}, i.e. for $\tau=0$, but for a general $\mu$ (similar to~\cite{Garnier}). For a general 
$\tau$, the main change which affects the evolution of the rarefaction wave is the speed of sound and its temporal dependence. 
Since $c_s(t)=c_0{\left(t/t_s\right)}^{\frac{\beta\tau}{2}}$ the rarefaction wave for $\tau>0$ travels at a lower speed of sound than $c_0$ in early times ($t<t_s$), and higher speed of sound than $c_0$ in later times ($t>t_s$). For $\tau<0$ it is the opposite: a higher speed of sound than $c_0$ in early times, and a lower one in late times. However, an exact solution of the rarefaction wave may be found only for $\tau=0$. To count this effect, we use the solution for $\tau=0$, using the $\tau$-dependent $c_0$ averaged in time. In a matter of fact, since in Eqs.~\ref{delta_kinetic}-\ref{delta_total}, we have the third power of the speed of sound, we average over ${c_s^3(t)}$ in time until time $t$:
\begin{equation} \label{E_time_fac}
\overline{c_s^3(t)}=\frac{\int_{0}^{t}c_s^3(t')dt'}{\int_{0}^{t}dt'}=\frac{\int_{0}^{t}c_0^3{\left(\frac{t'}{t_s}\right)}^{\frac{3\beta\tau}{2}}dt'}{t}=
\frac{c_0^3{\left(\frac{t}{t_s}\right)}^{\frac{3\beta\tau}{2}}}{1+3\beta\tau/2}
\end{equation}     
The average expression for $\overline{c_s^3(t)}$ is an approximation to count for the exact time evolution of the speed of sound and its effect on the rarefaction wave. It includes the temporal behavior $(t/t_s)^{\frac{3\beta\tau}{2}}$, and an additional pre-factor of $1/(1+3\beta\tau/2)$. Both of them reproduces~\citet{HR_PRE} for the case of $\tau=0$ yielding $\overline{c_s^3(t)}=c_0^3$. Under these assumptions, the final correction for the energy in the supersonic case, including the temporal dependence, is:
\begin{equation} \label{E_super_final}
\Delta{E_{\mathrm{Super,tot}}(t)}/A=\frac{2(\mu+r)}{r(1-\mu)(2-3\mu)}\frac{\rho_0{c_0}^3{\left(\frac{t}{t_s}\right)}^{\frac{3\beta\tau}{2}}}{1+3\beta\tau/2}t,
\end{equation}  
and the total energy stored in the matter in given by Eq.~\ref{final_super_cor}. In Sec.~\ref{sec:General Power-Law Boundary Condition for Gold} we show this simple averaging yields surprisingly a very good agreement with exact simulations for all values of $\tau$. Explicitly, the results for the total energy in the supersonic case for gold and for \TaO by Eq.~\ref{E_super_final} for $\tau=0$ are:
\begin{subequations}
\begin{equation} \label{E_super_final_gold}
E_{\mathrm{Au,Super,full}}/A=2.86{T_0}^{3.55}{\rho_0}^{-0.17}{t}^{0.5} +4.54{T_0}^{2.4}{\rho_0}^{0.79}{t}\quad\mathrm{[kJ/cm^2]}
\end{equation} 
\begin{equation} \label{E_super_final_Ta2O5}
E_{\mathrm{Ta_2O_5,Super,full}}/A=3.09{T_0}^{3.575}{\rho_0}^{-0.175}t^{0.5} +8.02{T_0}^{2.05}{\rho_0}^{0.83}{t}\quad\mathrm{[kJ/cm^2]}
\end{equation} 
\end{subequations}
The values for Au in Eq.~\ref{E_super_final_gold} are very similar to the values in Eq. (6) in~\cite{HR_PRE} with a slight difference of factor of $\frac{2\sqrt{1-\mu}(r+\mu)}{(2-3\mu)(r+\mu(1-\mu)^{3/2})}$ in the constant of the second term (the constant 4.54 in Eq.~\ref{E_super_final_gold}, compared to 3.5 in~\cite{HR_PRE}). This is due to the fact the Ref.~\cite{HR_PRE} assumes $\mu=0$ for the rarefaction wave solution, and thus uses the approximated Eq.~\ref{approx_rare}, while we used the more exact solution (general $\mu$), Eq.~\ref{exact_rare}, similarly to~\cite{Garnier}.

 
Similarly to the supersonic case, we apply a correction for the energy in the subsonic regime as well. For $\M>1$ and for reasonable values of $\tau$ ($\tau<\tau_c$), the Mach number decreases in time. Hence, the heat wave starts propagating as a supersonic wave, and turns subsonic at a certain time. We use the given rough estimation for this time, $t_{\mathrm{catch}}$ of~\cite{HR_PRE}, as the time when the rarefaction front takes over the heat front. Solving the equation $x_f(t)=x_r(t)=c_s(t)t$ for $t=t_{\mathrm{catch}}$ yields (assuming $\tau=0$, as we need only a rough estimation of $t_{\mathrm{catch}}$):
\begin{equation}  \label{t_catch}
t_{\mathrm{catch}}=\frac{16\sigma{g}}{3(4+\alpha)}\frac{{T_0}^{4+\alpha-2\beta}}{r(1-\mu)f^2{\rho_0}^{2+\lambda-2\mu}}\xi_{f,\mathrm{Super}}^2 
\end{equation}  
Up to a dimensionless constant of order of unity, this is the typical time $t_s$ for which $\chi_0$ from Garnier's solution is equal to 1~\cite{Garnier}.
Explicitly, the quantitative results for $t_{\mathrm{catch}}$ for gold and for \TaO are:
\begin{subequations}
\begin{equation} \label{tcat_gold}
t_{\mathrm{catch,Au}}=0.182{T_0}^{2.3}\rho_0^{-1.92}\quad\mathrm{[ns]}
\end{equation} 
\begin{equation} \label{tcat_Ta2O5}
t_{\mathrm{catch,Ta_2O_5}}=0.06{T_0}^{3.05}\rho_0^{-2.01}\quad\mathrm{[ns]}
\end{equation} 
\end{subequations}

The correction for the subsonic case is carried out by subtracting the energy from the pure subsonic energy term for $t<t_{\mathrm{catch}}$, and adding up instead the {\bf{full}} energy terms for the supersonic case (including the rarefaction supersonic correction), for which the heat wave was still supersonic. Hence, the correction is given by:
\begin{align} \label{E_sub_correction}
\Delta{E_{\mathrm{Sub,tot}}}/A=&-{E_{\mathrm{Sub,pure}}}(t_{\mathrm{catch}})/A+ \\ \nonumber
&E_{\mathrm{Super,pure}}(t_{\mathrm{catch}})/A+\Delta{E_{\mathrm{Super,tot}}(t_{\mathrm{catch}})}/A,
\end{align}
and the total energy stored in the matter in the subsonic regime is:
\begin{equation}
\label{final_sub_cor}
E_{\mathrm{Sub,tot}}/A=E_{\mathrm{Sub,pure}}/A+\Delta{E_{\mathrm{Sub,tot}}}/A,
\end{equation}
using $E_{\mathrm{Sub,pure}}/A$ from Eq.~\ref{E_sub_1}, $E_{\mathrm{Super,pure}}/A$ from Eq.~\ref{E_super_1} and $\Delta{E_{\mathrm{Super,tot}}}/A$ from Eq.~\ref{final_super_cor}.
Explicitly, the results for the total energy in the subsonic case for gold and for \TaO, including the correction from Eqs.~\ref{E_sub_correction} for $\tau=0$ are:
\begin{subequations}
\begin{equation} \label{E_sub_final_gold}
E_{\mathrm{Au,Sub,full}}/A=5.92{T_0}^{3.35}t^{0.59} -0.12{T_0}^{4.7}{\rho_0}^{-1.13}\quad\mathrm{[kJ/cm^2]}
\end{equation} 
\begin{equation} \label{E_sub_final_Ta2O5}
E_{\mathrm{Ta_2O_5,Sub,full}}/A=7.21{T_0}^{3.31}t^{0.59} -0.133{T_0}^{5.1}{\rho_0}^{-1.18}\quad\mathrm{[kJ/cm^2]}\end{equation} 
\end{subequations}
The values for Au in Eq.~\ref{E_sub_final_gold} are very similar to the values in Eq. (7) in~\cite{HR_PRE} with a slight differences due to the general $\mu$ modifications.
\begin{figure}[htp]
	\centering
	\includegraphics[width=7.5cm]{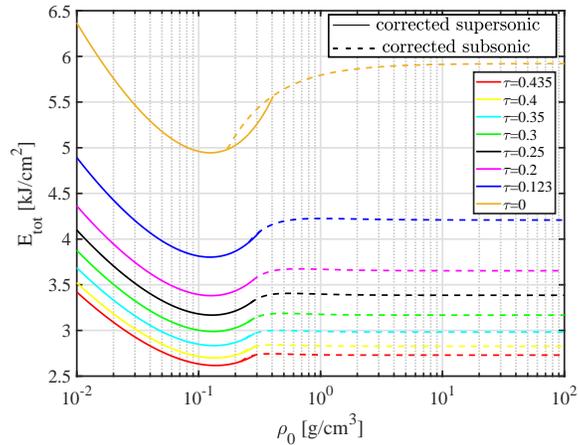}
	\caption{The approximated supersonic and subsonic expressions for the total energy per area unit versus the initial density $\rho_0$ for various $\tau$ values for gold, $T_0=100 \mathrm{eV}$ and $t=1 \mathrm{nsec}$.}
	\label{fig:E_vs_rho0_Mordy_all_taus}
\end{figure}

In Fig.~\ref{fig:E_vs_rho0_Mordy_all_taus} we present a demonstration of the total energy per area unit as a function of the initial density $\rho_0$ {\em{for several values of}} $\tau$, within the range $0\leqslant\tau\leqslant\tau_c$ for gold, for temperature of $T_0=100 \mathrm{eV}$ and time of $t=1 \mathrm{nsec}$. For all cases, the energy was calculated from the expressions for ${E_{\mathrm{Super,tot}}}(t)/A$ and ${E_{\mathrm{Sub,tot}}}(t)/A$, including the correction terms. The curve for each $\tau$ contains two branches - for the supersonic (solid lines) and subsonic (dashed lines) regimes, respectively. We can see that the energy has a minimum point at some optimal density $\rho_{0,\mathrm{min}}$, when the optimal density is always in the supersonic branch; there is a clear minimum point in the supersonic branch at $\rho_{0,min}\approx0.125 \mathrm{g/cm^3}$ for all values of $\tau$. A detailed comparison of these approximations with the extended Garnier's self-similar solution results and with numerical simulations results is presented in the next section.   

\section{Main results} 
\label{sec:Main results}

In this section, we present the main results which obtained using the physical models described in the previous section. We present an examination for the extension of the critical case for gold, while checking the validity range of this extension in terms of $\tau$. The sensitivity to the matter properties is examined, in particular, the EOS, and we compare between an analytic EOS and a SESAME tabular EOS for gold. The analysis of the transition region is obtained also for \TaO.~All of these results are tested and compared against numerical simulation results. As discussed, we focus on those materials, which are commonly used in similar studies (usually for hohlraum experiments)~\cite{HR_PRE,HR,exp_PRL}.

The simulations were performed using a one-dimensional radiative-hydrodynamics code, which couples Lagrangian hydrodynamics with implicit LTE diffusion radiative conduction scheme (first and second order of discretization, for time and space, respectively), in a first order operator-split method. The code uses explicit hydrodynamics using Richtmyer's artificial viscosity (include both linear and squared terms to reduce numerical instabilities near shock fronts) and Courant's criterion for a time-step. In the diffusion conduction scheme, the time-step is defined dynamically such that the temperature will not change in each cell by more than 5\% between time steps (for more details regarding the radiative-hydrodynamics code, see~\cite{shussman,binary_EOS,avner1,avner2,alum}). In all simulations, we have used a converged constant space intervals. We have performed numerical simulations for a large variety of parameters: $\rho_0$, $T_0$, $\tau$, correspond to a wide range of $\chi_0$ and $\M$ values, for different cases: a gold (Au) slab with an analytic EOS as in Eq.~\ref{pwrlaws}, a gold slab while assuming an exact SESAME tabular EOS and a \TaO slab with an analytic EOS. The hydrodynamic profiles in this section are presented as physical quantities with dimensions and plotted as a function of the Lagrangian mass coordinate. One can obtain the dimensionless profiles form using Eqs.~\ref{pwrlaws},~\ref{Ans_Garnier}. The value of $\chi_0$ is presented for each profile. Several figures show relevant physical quantities as a function of $\rho_0$, similarly to previous studies (one can easily transform between $\rho_0$ to dimensionless parameters such as $\chi_0$ or $\M$ using Eqs.~\ref{Mach_super},~\ref{chi0},~\ref{Mach_super_G}).

\subsection{General Power-Law Boundary Condition for Gold}
\label{sec:General Power-Law Boundary Condition for Gold}

In this section we examine the extension of the self-similar solution for the critical case to a general $\tau$.
This examination allows us to evaluate the validity range of $\tau$ values, for which the extended solutions are a good approximation, compared to the exact solution, which is obtained via numerical simulation. In Sec.~\ref{garnier_extension} this validity range was evaluated to be for $\tau\gtrapprox0.25$ (for gold). The values of $\tau$ we examine in this section are in the range of $0\leqslant\tau\le\tau_c$. We first show the results for the self-similar parameters, obtained from the extended self-similar solver. Then we examine the solutions obtained from the solver by looking at the hydrodynamic profiles. Lastly, we characterize the transition region between the supersonic and subsonic regimes, using physical quantities whose sensitivity in this region is significant, such as: the energy as a function of the initial density, the compression ratio of the compression/shock wave and the maximal value of the pressure profile.

\begin{figure}[htp]
	\centering
	(a)
	\includegraphics[width=7.5cm,clip=true]{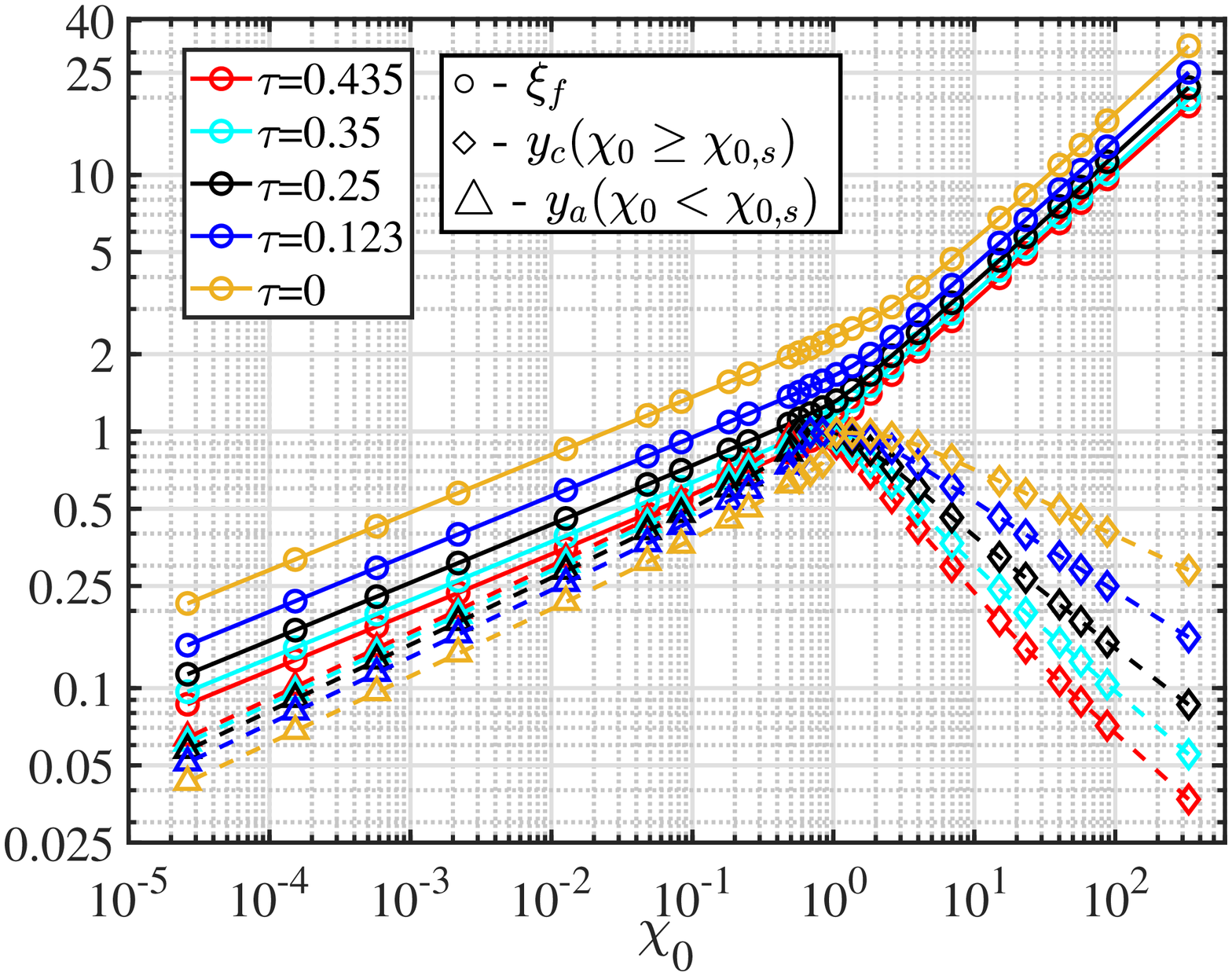}
    (b)
	\includegraphics[width=7.4cm,clip=true]{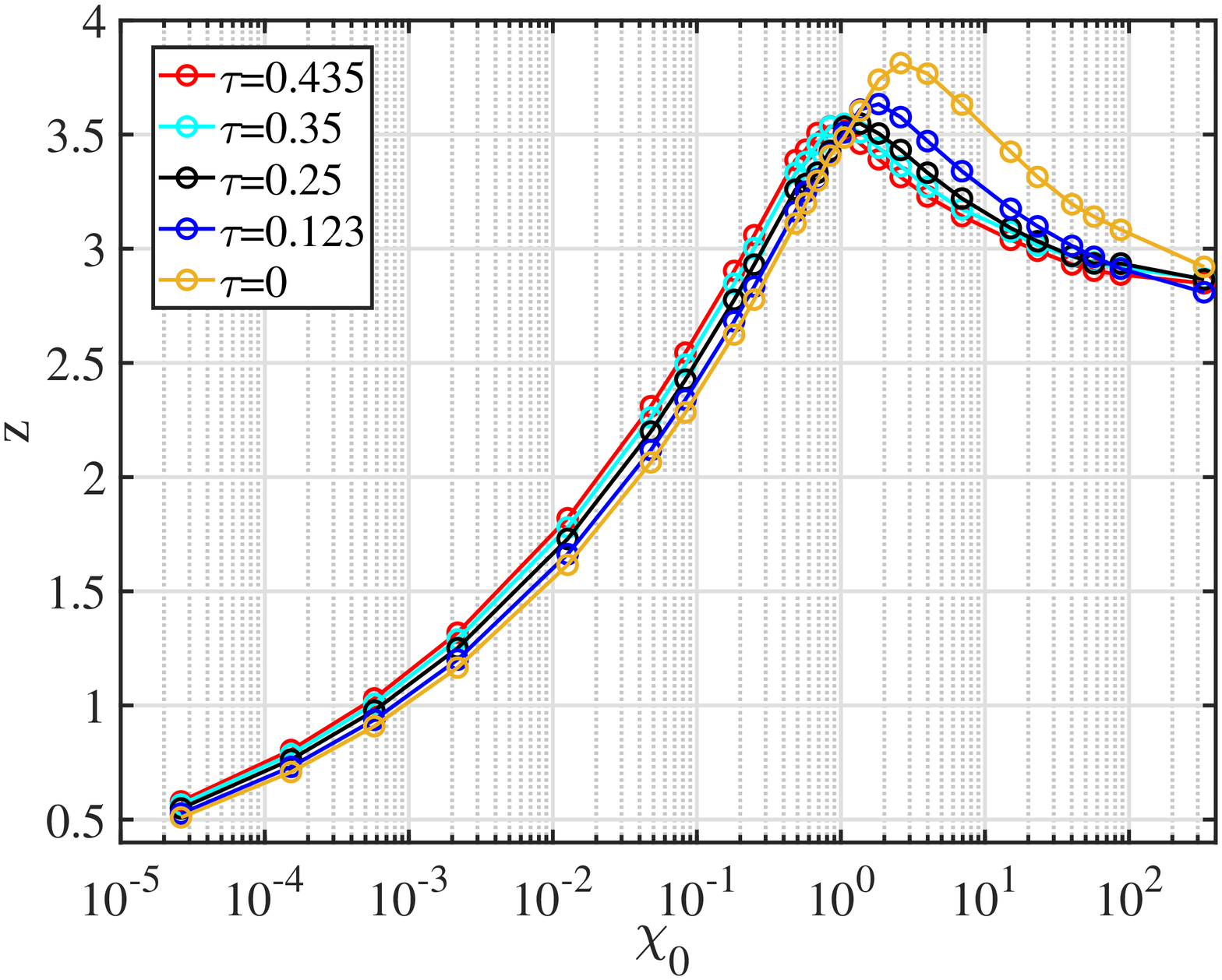}
	\caption{(a) The dimensionless parameters of the self-similar solution for gold as a function of $\chi_0$ for various values of $\tau$. $\chi_{0,s}$ is the cut off parameter between a solution with $y_c$ to a solution with $y_a$. $\chi_{0,s}=1$ for $\tau=0$, $\chi_{0,s}=0.6$ for $\tau=0.123$, and $\chi_{0,s}=0.5$ for all other values of $\tau$. (b) The dimensionless total energy of the self-similar solution $\ti{z}$ for gold as a function of $\chi_0$ for various values of $\tau$.}
	\label{fig:z_vs_chi0_for_taus_Au}
\end{figure}

In Fig.~\ref{fig:z_vs_chi0_for_taus_Au}(a) the self-similar parameters $\xi_f$ and $y_c,y_a$ are presented as a function of $\chi_0$ for various $\tau$ values. It can be seen that for any $\tau$, at some value of $\chi_0$ ($\approx 0.5$ in our case), the solver's solution method is changed from finding $y_c$ (the self-similar isothermal shock front) to finding $y_a$ (the self-similar ablation front). This occurs by definition at $\chi_0$ for which $y_c\approx1$, indicating the heat wave is in the deep subsonic regime, and the shock wave overtakes the ablation heat front. We can see that in the supersonic regime $y_c$ starts close to 0 for the highest value of $\chi_0$, indicating the heat front is much ahead of the isothermal shock front. As $\chi_0$ decreases, $y_c$ increases, since the shock comes closer to the heat front. For $\chi_0<0.5$ the shock wave is ahead of the ablation front, and as $\chi_0$ decreases, so does the Mach number, and the heat wave becomes more subsonic. Since $y=1$ indicates the shock front location for this branch of the solver, the decrease of the Mach number can be seen through the decreasing value of $y_a$, meaning the relative distance between the heat front and the shock front increases.

For a given value of $\chi_0$, $\xi_f$ and $y_c$ decrease with $\tau$, as $y_a$ increases with $\tau$. These trends indicate that as $\tau$ decreases, the obtained solution for a given $\chi_0$ has a lower Mach number - in the supersonic region the isothermal shock propagates faster (thus, $y_c$ is higher), and in the subsonic region the heat front is even farther behind the shock wave (thus $y_a$ is lower).  

Another important set of dimensionless parameters are the dimensionless energies: $\ti{z}_{k}$, $\ti{z}_{int}$, and $\ti{z}$, for the dimensionless kinetic energy, the dimensionless internal energy and the dimensionless total energy (where $\ti{z}=\ti{z}_{k}+\ti{z}_{int}$). Their values are determined by the following integrals over the shape functions:
$\ti{z}_{k}=\int_{0}^{1}{\frac{1}{2}\xi_f^2{\ti{u}}^2dy}$,
$\ti{z}_{int}=\int_{0}^{1}{\frac{1}{r}{\ti{V}}^{\mu}{\ti{\zeta}}^{\varepsilon}dy}=\int_{0}^{1}{\frac{1}{r}{\ti{V}}^{\mu}{\ti{T}}^{\beta}dy}$.
To obtain the physical energies (all of the energies are per area unit), those dimensionless parameters are multiplied by: $E_1=rf{T_0}^\beta{\rho_0}^{-\mu}m_{f0}{\left(\frac{t}{t_s}\right)}^{n+2j}={\xi_f}(rf{T_0}^\beta{\rho_0}^{-\mu})^{\nicefrac{3}{2}}{\rho_0}t_s{\left(\frac{t}{t_s}\right)}^{3n+2\ell-2}$. The dimensionless total energies are presented in Fig.~\ref{fig:z_vs_chi0_for_taus_Au}(b) as a function of $\chi_0$ for various $\tau$ values. The dimensionless energies have a maximum near $\chi_0=1$, which is close to the sonic point. In the extreme supersonic regime for high values of $\chi_0$, the internal energy is high, but there is almost no kinetic energy. As $\chi_0$ decreases, the fraction of $\ti{z}_{k}/\ti{z}$ increases, indicating the importance of the hydrodynamic motion for $\M\lessapprox1$. The transition to the subsonic regime starts when this ratio is about $0.1-0.15$. This ratio reaches an asymptotic value of $\sim{0.25}$ when $\chi_0\to 0$. In the subsonic regime, the dimensionless energies decrease with $\chi_0$, since the ablation area is getting relatively smaller ($y_a$ decreases as mentioned). The heat wave region contains much more energy compared to the shock region, mainly because the temperatures behind the heat front are much higher. The dimensionless total energy is relatively a weak function of $\tau$ (for $\tau>0.1$); only the $\tau=0$ curve seems to be differ significantly.

\begin{figure}[htp]
	\centering
	\includegraphics[width=16cm]{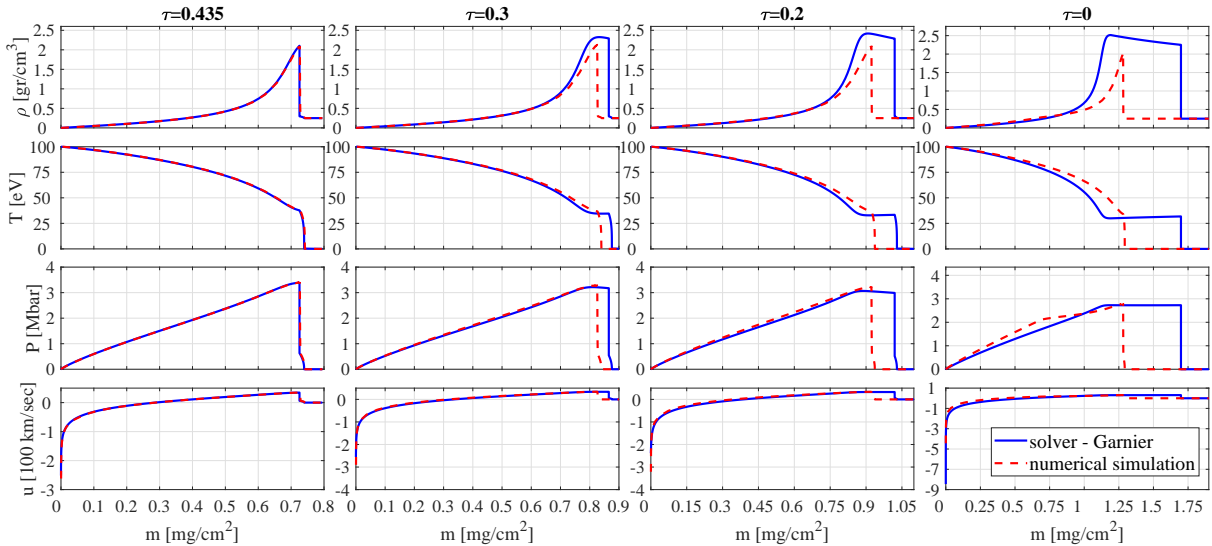}
	\caption{The hydrodynamic spatial profiles for gold at $\chi_0=0.69$: $\rho_0=0.25 \mathrm{g/cm^3}$, $T_0=100$eV, $t=1$nsec for various values of $\tau$.}
	\label{fig:profiles_Au_all_taus_T0=100_t=1}
\end{figure}

Next, the hydrodynamic profiles from the self-similar solver for $\rho_0=0.25 \mathrm{g/cm^3}$, $T_0=100 \mathrm{eV}$ and $t=1 \mathrm{nsec}$ ($\chi_0=0.69$) are presented in Fig.~\ref{fig:profiles_Au_all_taus_T0=100_t=1} for four values of $\tau$: $\tau_c=0.435$, $0.3$, $0.2$, and $0$. In all four cases, the results from the solver are compared against 1D numerical simulation results. The first case for $\tau=\tau_c$ is presented as a point of reference when the matching between the solver and the simulations is perfect (for other values of $\chi_0$ for $\tau=\tau_c$, see Appendix B). For the other cases, we can see for any of the hydrodynamic functions, that as $\tau$ decreases, the agreement between the extended self-similar solution to the numerical simulation results becomes worse.

\begin{figure}
    (a)
		\includegraphics[width=7.5cm,clip=true]{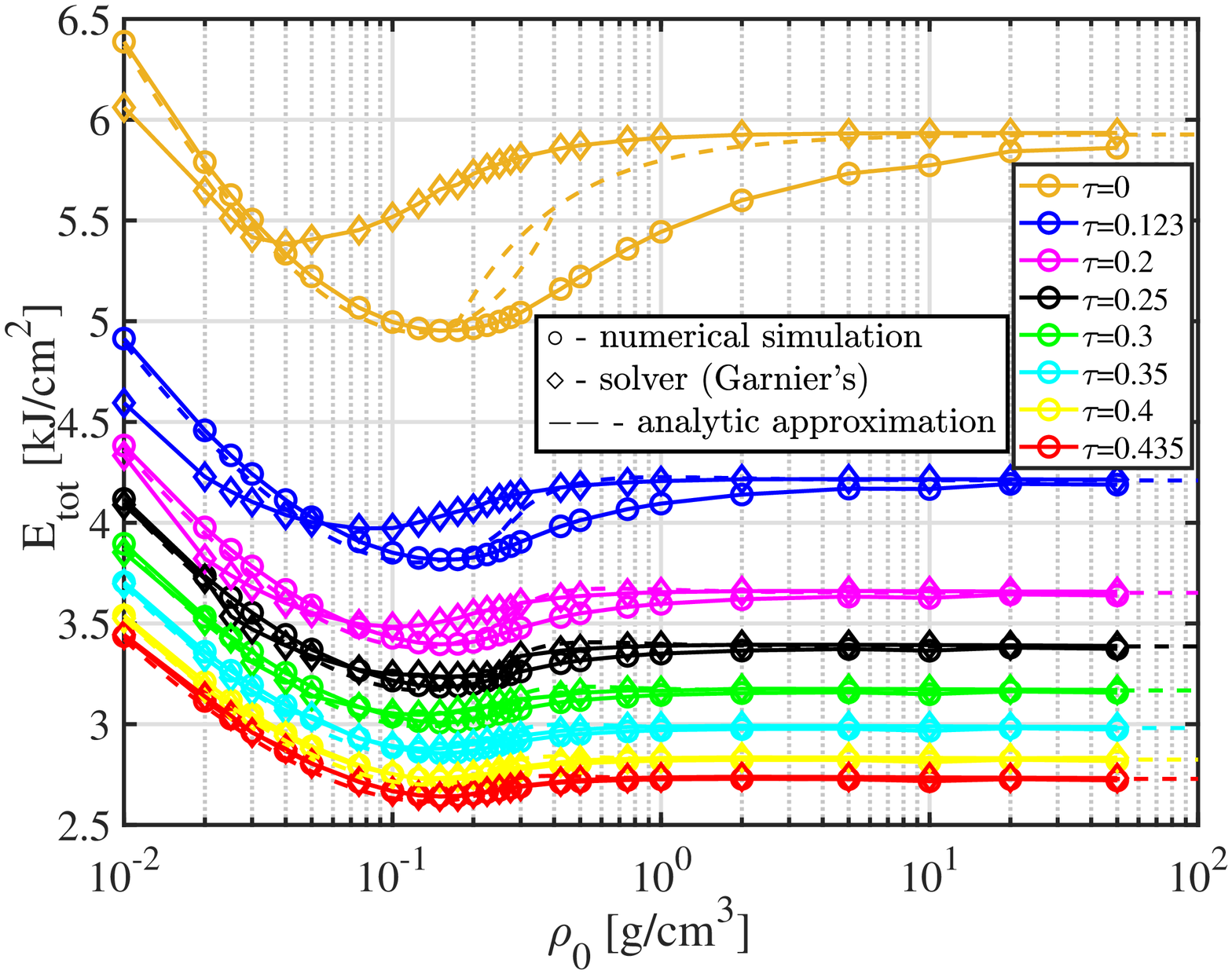}
	(b)
		\vspace{5mm}
		\includegraphics[width=7.5cm,clip=true]{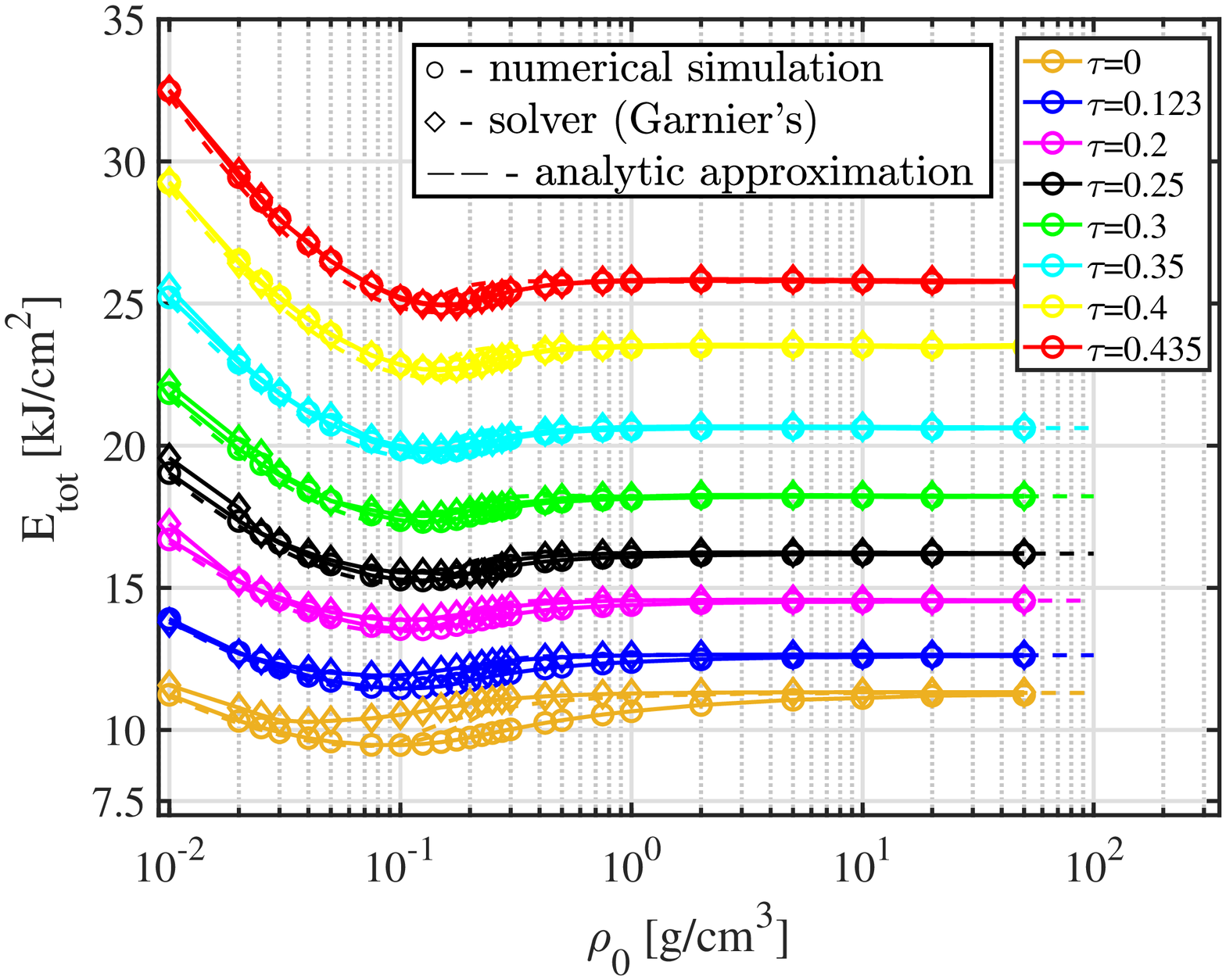}
	(c)
    \includegraphics[width=7.5cm,clip=true]{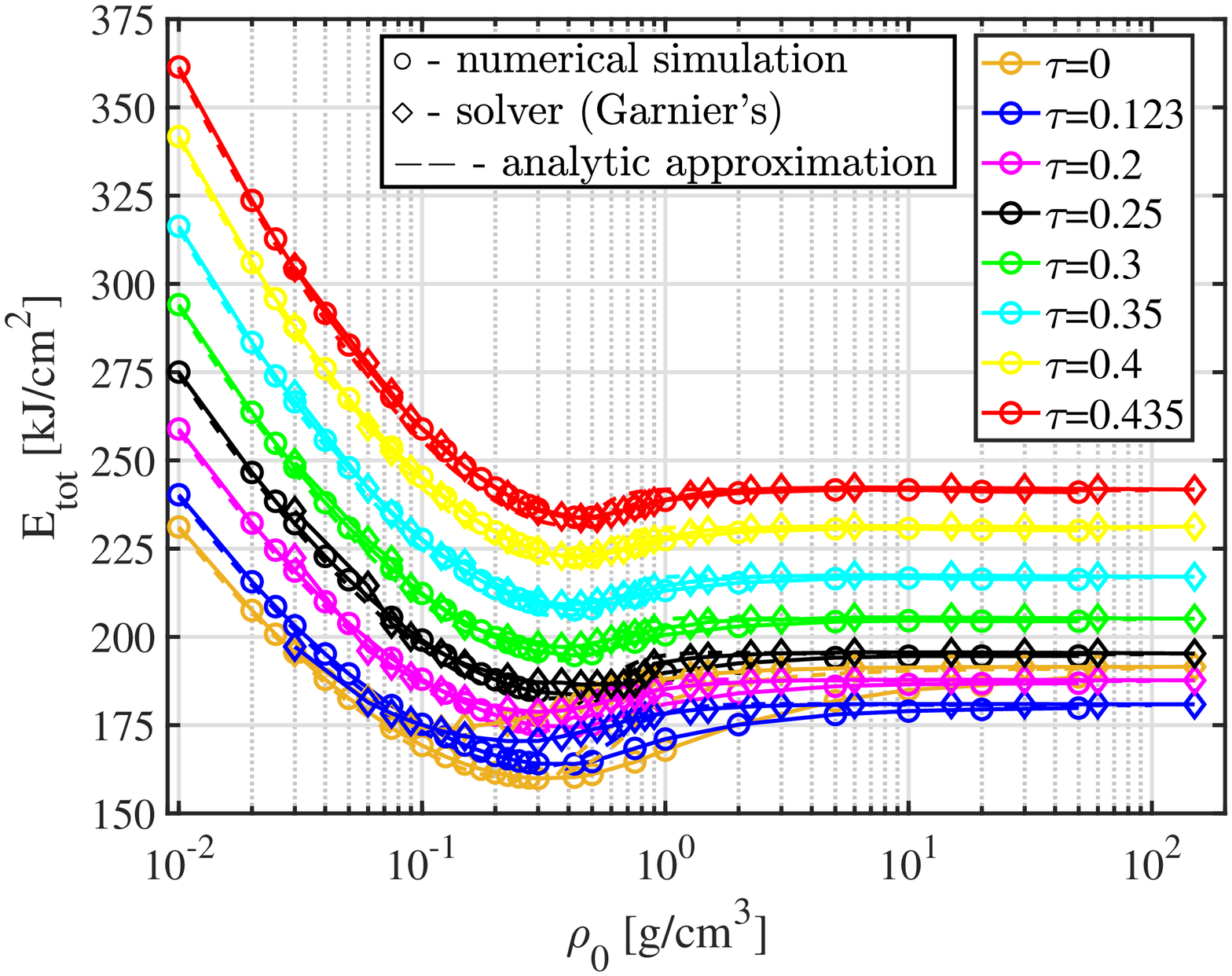}	
		\caption{The energy per area unit Vs. the initial density $\rho_0$ for gold at: (a) $T_0=100$eV, $t=1$nsec, (b) $T_0=100$eV, $t=3$nsec, (c) $T_0=250$eV, $t=2$nsec, for various values of $\tau$. } \label{fig:E_tot_vs_rho0_all_taus_Au_T0=100eV_t=3nsec}
\end{figure}

Now, we characterize the transition region between the supersonic and subsonic branches, using physical quantities whose sensitivity in this region is significant. We start by examination of the energy (per area unit) as a function of the initial density, for several cases (as in~\cite{HR_PRE}): for $T_0=100 \mathrm{eV}$ and $t=1 \mathrm{nsec}$ in Fig.~\ref{fig:E_tot_vs_rho0_all_taus_Au_T0=100eV_t=3nsec}(a), for $T_0=100 \mathrm{eV}$ and $t=3 \mathrm{nsec}$ in Fig.~\ref{fig:E_tot_vs_rho0_all_taus_Au_T0=100eV_t=3nsec}(b), and for $T_0=250 \mathrm{eV}$ and $t=2 \mathrm{nsec}$ in Fig.~\ref{fig:E_tot_vs_rho0_all_taus_Au_T0=100eV_t=3nsec}(c). In all three cases we compare the total energy which was obtained from the self-similar solver presented in Sec.~\ref{garnier_extension} (in diamonds) to the total energy from the numerical simulation (in circles) and to the total energy from the analytic approximation presented in Sec.~\ref{approximated} (dashed curves).
   
Since the problem is self-similar (for $\tau=\tau_c$) it is not surprising that the shape of the curve stays similar (qualitatively) for all the different cases. In all these figures, we can see a decrease of the energy in the supersonic regime as $\rho_0$ increases, until the minimal value is reached at $\rho_{0,\min}$ (within the supersonic regime), and then as $\rho_0$ keeps increasing, the total energy increases as well, indicating the heat wave is in the transition region. For high enough values of $\rho_0$ the total energy approaches its asymptotic value in the deep subsonic regime. 

The sonic point itself is obtained at $\rho_0\approx0.2 \mathrm{g/cm^3}$ in the first and second cases (Figs.~\ref{fig:E_tot_vs_rho0_all_taus_Au_T0=100eV_t=3nsec}(a) and ~\ref{fig:E_tot_vs_rho0_all_taus_Au_T0=100eV_t=3nsec}(b)), and at $\rho_0\approx0.7 \mathrm{g/cm^3}$ in the third case (Fig.~\ref{fig:E_tot_vs_rho0_all_taus_Au_T0=100eV_t=3nsec}(c)). It is worth mentioning that although the total energy for a given initial density depends on $\tau$ significantly, the transition region in terms of $\rho_0$ values, and the sonic point in particular, have a quite low sensitivity for $\tau$ (for given $T_0$ and time). Since the Mach number stays constant in time for $\tau=\tau_c$ (and approximately for the other $\tau$ values, as well), the sonic point does not change in time (between $t=1 \mathrm{nsec}$ and $t=3 \mathrm{nsec}$ in our case). That is because $\chi_0$ does not depend on $t$. However, it depends on the boundary temperature $T_0$, and hence, there is a change in the sonic point in terms of $\rho_0$ for $T_0=250 \mathrm{eV}$ in the third case compared to the first two.

The results from the analytic approximation are in a good agreement with the simulations for any value of $\tau$ (better than $1.5\%$ in the far supersonic and subsonic branches, and better than $5\%$ in the transition region, where the two branches are patched). Next, it is evident from Fig.~\ref{fig:E_tot_vs_rho0_all_taus_Au_T0=100eV_t=3nsec}(a) that for $\tau\geqslant0.25$ there is a good agreement between the extended self-similar solution, the simulation, and the analytic approximation.
For $\tau\leqslant0.2$ there is a difference, which grows as $\tau$ decreases, between the extended solution and the simulation results. For those values, the analytic approximation provides a better estimation for the total energy, where it yields a good agreement with the simulation results, especially in the supersonic regime. This range of values for $\tau$ is very close to the expected validity range evaluated in Sec.\ref{garnier_extension}.

The sensitivity of the solution for low $\tau$ values decreases for later times, as can be seen from Fig.~\ref{fig:E_tot_vs_rho0_all_taus_Au_T0=100eV_t=3nsec}(b). In this case we can see a good agreement between the solver and the simulation results, down to $\tau=0.123$. Again, the analytic approximation yields a good agreement for all values of $\tau$. For $t\leqslant1 \mathrm{nsec}$ the total energy for a lower $\tau$ is higher, because for $\tau_1<\tau_2$ and $t\leqslant{t_s}=1 \mathrm{nsec}$, $T_s(t,\tau_1)>T_s(t,\tau_2)$, and more energy enters the system for lower $\tau$ values. The opposite is true for later times: for $t\geqslant{t_s}=1 \mathrm{nsec}$, $T_s(t,\tau_1)<T_s(t,\tau_2)$, thus after enough time, the total energy for the higher $\tau$ will overcome the energy for the lower one. The third case presented in Fig.~\ref{fig:E_tot_vs_rho0_all_taus_Au_T0=100eV_t=3nsec}(c) shows the same qualitative picture. In that case, we can see a good agreement between the results of the solver and the simulation for $\tau\geqslant0.2$ and a good agreement between the simulation and the analytic approximation for all values of $\tau$.


Another parameter that can be used to characterize the transition region is the compression ratio of the compression/shock wave. For a given boundary temperature $T_0$ and for a given time $t$, we define for each value of $\rho_0$ the maximal density $\rho_{\max}$, which is the density at the shock front (or the rarefaction front in case it is too weak to be considered as a shock wave). Then, the dimensionless ratio $\rho_{\max}/\rho_0$ is examined as a function of $\rho_{0}$. In the deep supersonic regime, the compression is almost negligible, due to the negligence of the hydrodynamic motion, therefore, for very low initial densities we would expect $\rho_{\max}/\rho_0\approx1$. As the initial density increases, and the Mach number decrease, the hydrodynamic perturbation gets stronger, along with the compression behind the isothermal shock wave, hence, $\rho_{\max}/\rho_0$ increases with $\rho_0$. Since we assumed an EOS of a quasi-ideal-gas, the compression ratio behind the shock wave is bound by the maximal ratio in the limit of a strong shock. From Rankine-Hugoniot relations, this ratio is $\frac{\rho_{\max}}{\rho_0}=\frac{\gamma+1}{\gamma-1}=\frac{r+2}{r}$, where $\gamma\equiv{1+r}$ is the adiabatic index of the matter~\cite{zeldovich,shussman}. Thus, we expect the compression ratio to approach this asymptotic value when the heat wave is in the deep subsonic regime. The transition region is characterized by the monotonous increase of the compression ratio.

\begin{figure}[htp]
	\centering
(a)
\includegraphics[width=7.5cm,clip=true]{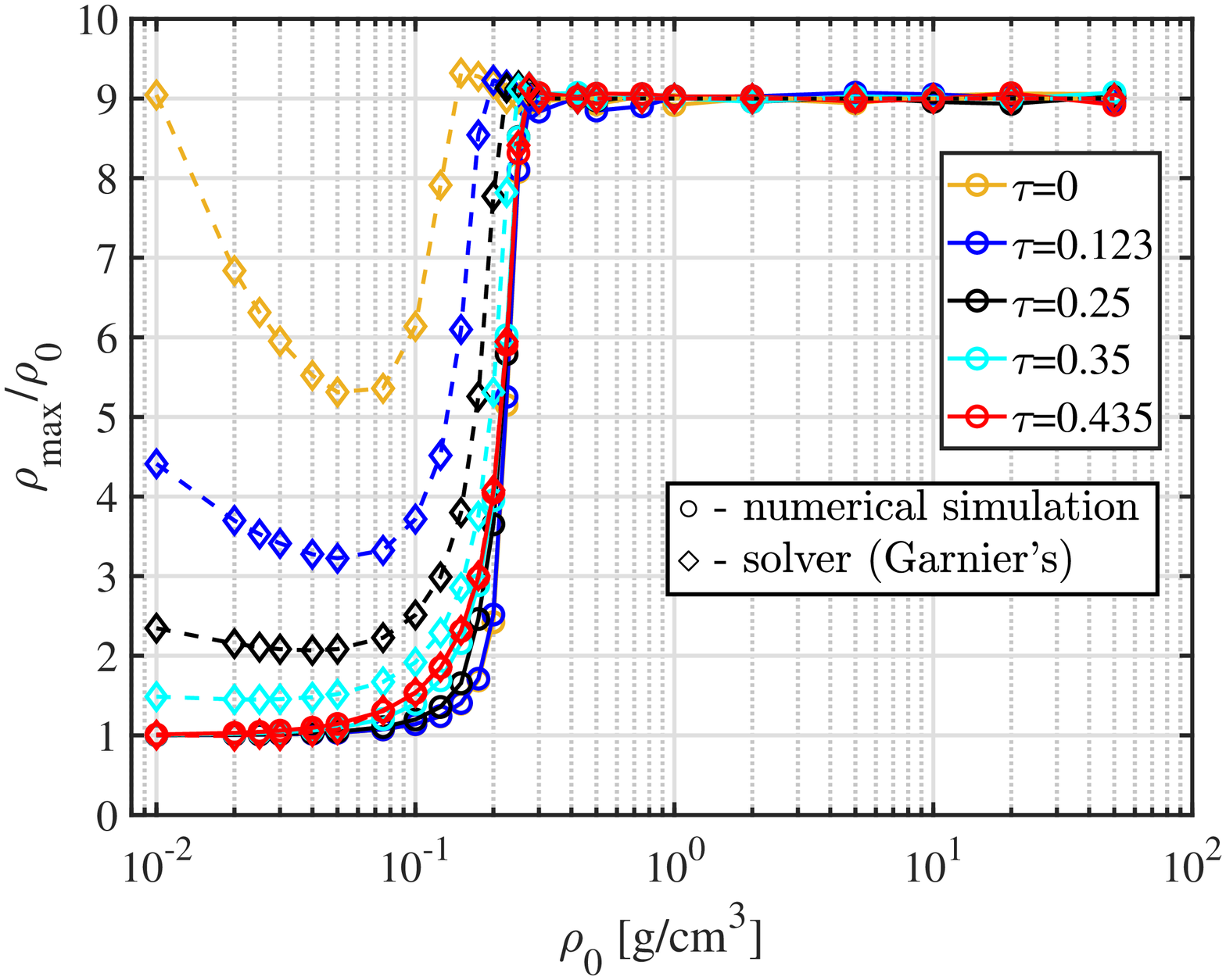}
(b)
\includegraphics[width=7.5cm,clip=true]{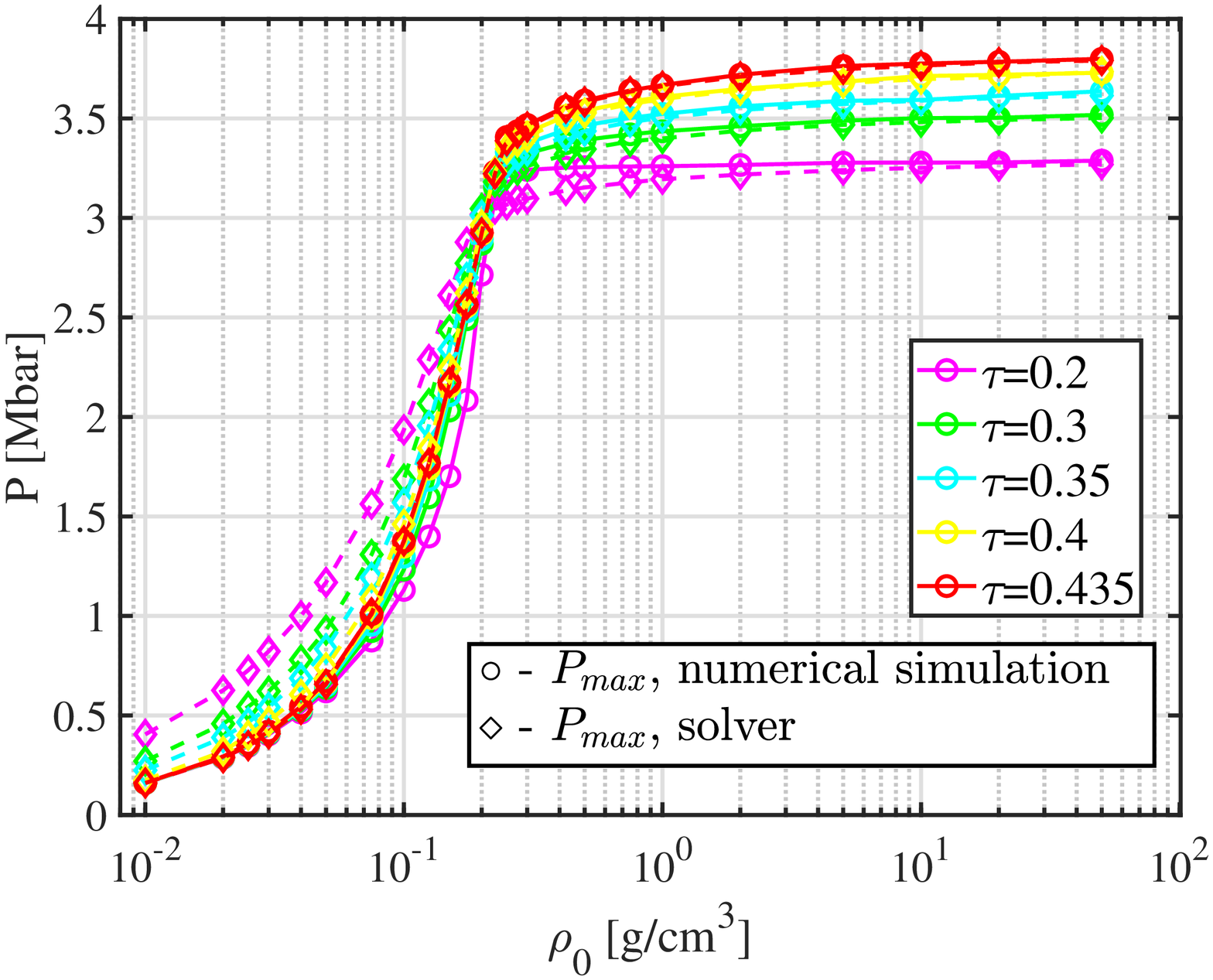}
\caption{(a) The ratio $\rho_{\max}/\rho_0$ Vs. the initial density $\rho_0$ for gold at $T_0=100$eV, $t=1nsec$ for various values of $\tau$. (b) The maximal pressure $P_{\max}$ Vs. the initial density $\rho_0$ for gold at $T_0=100$eV, $t=1$nsec for various values of $\tau$.}
\label{fig:rho_max_over_rho0_vs_rho0_all_taus_Au_T0=100eV_t=1nsec}
\end{figure}

Fig.~\ref{fig:rho_max_over_rho0_vs_rho0_all_taus_Au_T0=100eV_t=1nsec}(a) presents the compression ratio $\rho_{\max}/\rho_0$ as a function of $\rho_0$ for $T_0=100 \mathrm{eV}$ for several values of $\tau$. $\rho_{\max}/\rho_0$ yields the expected curve: the compression ratio approaches the asymptotic values of $1$ and $\frac{r+2}{r}=9$ (for $r=0.25$ for gold) in the extreme supersonic and subsonic regimes, respectively, and between them, in the transition region in particular, the ratio monotonically increases with $\rho_0$. As was mentioned for the energy, the sonic point in this case is near $\rho_0\approx0.2 \mathrm{g/cm^3}$. Then, the curve rises sharply, towards the maximal asymptotic value. Thus, this physical quantity serves as a good characterization for the transition region of the heat wave (like the previous works offered~\cite{hoarty1,hoarty2,hoarty3,french_new}). The extended solver yields good agreement with the simulations for $\tau>0.35$. Hence, this quantity is even more sensitive than the total energy, since a good agreement between the solver and the simulations seems to be obtained only for higher $\tau$'s. Similarly to the total energy, the compression ratio shows a low sensitivity for the value of $\tau$ in the numerical simulation results. 

In addition, we examine the pressure as another physical quantity, $P_{\max}$ - the maximal value of the pressure profile. By looking at the pressure profiles in Fig.~\ref{fig:profiles_Au_all_taus_T0=100_t=1} we can see that maximal pressure is monotonically increasing, as $\rho_0$ increases (and the Mach number decreases). In the deep subsonic regime, it is evident from Fig.~\ref{fig:profiles_Au_all_taus_T0=100_t=1} that the maximal pressure is obtained at the ablation front, and it approaches an asymptotic value. In the supersonic region, since the heat wave is ahead from the shock, the density at the heat front is in its initial value, while the temperature approaches $0$. Therefore, the pressure at the heat front approaches $0$ as well (see Eq.~\ref{pwrlaws} for the EOS).

In Fig.~\ref{fig:rho_max_over_rho0_vs_rho0_all_taus_Au_T0=100eV_t=1nsec}(b), we examine the sensitivity to the boundary condition ($\tau$) by looking at the maximal pressure ($P_{\max}$). We can see $P_{max}$ is continuously and monotonically increasing, and reaches an asymptotic value in the deep subsonic regime. The sonic point is obtained near the inflection point of the curve (where the change in the curve's slope is maximal). We see a good agreement between the solver and the simulations in the subsonic region, and a relatively good agreement for $\tau\ge0.35$ in the supersonic and the transition regions. Moreover, by looking in Fig.~\ref{fig:rho_max_over_rho0_vs_rho0_all_taus_Au_T0=100eV_t=1nsec}(b) at the values of $P_{\max}$ for different values of $\tau$, and the values of $\rho_{0}$ these jumps occur at, we deduce that the sonic point itself is less sensitive to the changes in $\tau$. Similarly to the compression ratio for the density, those pressure values can be used to characterize the transition region for the heat wave, and to quantify where this transition occurs.

\subsection{The Sensitivity to the EOS in Gold}  \label{sec:sesame_results}
In this section, we examine the sensitivity of the solution to the EOS of the matter, by comparing previous results with the results of numerical simulations, using a SESAME tabular EOS~\cite{SESAME}. The analytic EOS with the fitted parameters in table~\ref{table:pwr_law_opac_eos_Au} were taken from the EOS in~\cite{HR} that was fitted to 100-200eV regime, but is not a good fit for temperatures lower than 100eV, comparing to the exact SESAME EOS (see~\cite{binary_EOS}). This is relevant mainly in the shock region in the subsonic case, as can be seen from the temperature profiles in Fig.~\ref{fig:profiles_Au_all_taus_T0=100_t=1}. From the work of~\citet{binary_EOS,alum} we know that the shock front, and the hydrodynamic profiles in that region are sensitive to the EOS. Thus, we expect to see a significant sensitivity in the {\em{deep subsonic}} region. The examination presented in this section is meant to check whether the transition region is sensitive to the EOS in the same manner.

\begin{figure}[htp]
	\centering
	\includegraphics[width=16cm]{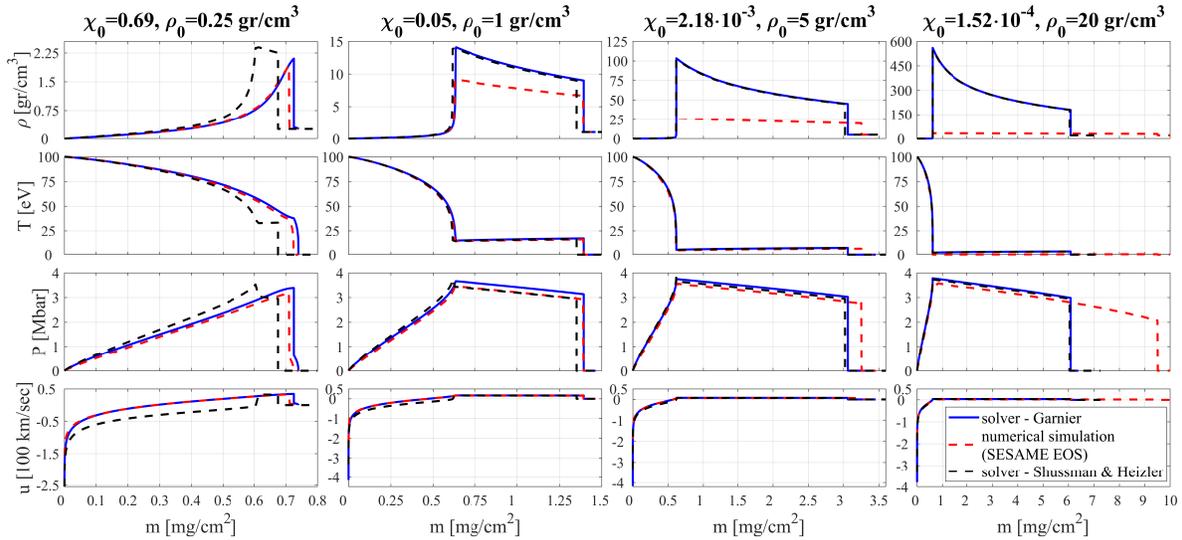}
	\caption{The hydrodynamic spatial profiles for various initial densities $\rho_0$ (and $\chi_0$) for gold at $T_0=100$eV, $t=1$nsec in the critical case ($\tau=\tau_c$). A comparison between an analytic EOS to SESAME EOS for different values of $\rho_0$.}
	\label{fig:profiles_Au_sesame_high_density_T0=100_t=1}
\end{figure}

The hydrodynamic profiles from the self-similar solver for $\tau=\tau_c$, $T_0=100 \mathrm{eV}$ and $t=1 \mathrm{nsec}$ ($\chi_0=0.69$) are presented in Fig.~\ref{fig:profiles_Au_sesame_high_density_T0=100_t=1} for some values of $\rho_0$ and $\chi_0$ in both the transition region and the subsonic region. 
The solver results are compared with both the self-similar solution obtained from~\cite{shussman}, and against numerical simulation results, using SESAME EOS. We can see a quite good agreement between the solver results and the simulation in the transition region (Shussman's et al. approximation is less good, since its validity is in the deep subsonic, strong shock regime). 
While the value of $\chi_0$ decreases ($\rho_0$ increases), the heat wave becomes deeper subsonic, so of course, we can see an excellent agreement between the solver and Shussman's self-similar approximation. 
However, a large deviation can be seen in the shock region as $\chi_0$ decreases, comparing to the exact numerical simulations with the exact SESAME EOS, especially in the density profile and the shock front position.

\begin{figure}
		\centering
		\includegraphics[width=7.5cm]{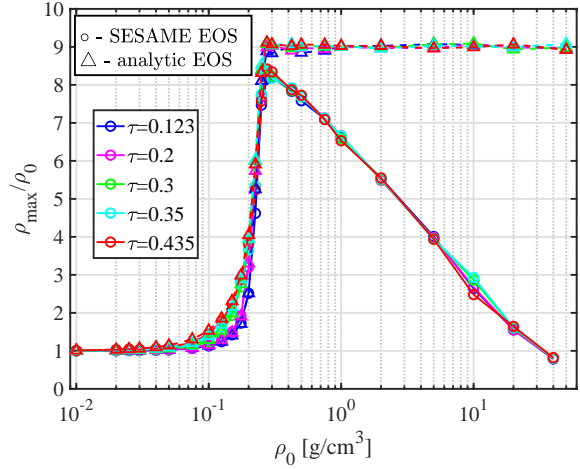}\\
		\caption{The ratio $\rho_{\max}/\rho_0$ vs. the $\rho_0$ for gold at $T_0=100$eV, $t=1$nsec for various values of $\tau$. A comparison between numerical simulations using an analytic EOS and SESAME EOS.}
		\label{fig:rho_max_over_rho0_vs_rho0_all_taus_Au_T0=100eV_t=1nsec_sesame}
\end{figure}

Next, we examine quantitatively the sensitivity to the EOS by looking at the compression ratio $\rho_{\max}/\rho_0$ as a function of $\rho_{0}$. The compression ratio is presented in Fig.~\ref{fig:rho_max_over_rho0_vs_rho0_all_taus_Au_T0=100eV_t=1nsec_sesame}, for  $T_0=100 \mathrm{eV}$ and $t=1 \mathrm{nsec}$, for various values of $\tau$. All values of $\tau$ show a similar behavior of the SESAME EOS simulations results, compared to the analytic EOS simulations results. Except for the strong shock regime in the subsonic region, there is an excellent agreement between the analytic EOS results and the SESAME EOS results in both the supersonic region and the transition region.

As explained, there is a significant sensitivity to the EOS in the deep subsonic region, as we have seen from the hydrodynamic profiles (Fig.~\ref{fig:profiles_Au_sesame_high_density_T0=100_t=1}). Since the analytic EOS fit is not a good description for those low temperatures ($<100$eV), the largest deviations occur at the shock region, especially for the density. Therefore, as $\rho_{0}$ increases, the deviation between the compression ratio from the analytic EOS and from the SESAME EOS simulations grows significantly. Nevertheless, it is worth mentioning that the compression ratio from both simulations (analytic EOS and SESAME EOS) shows a very low sensitivity to the temporal boundary condition, i.e., to the value of $\tau$, for all Mach numbers - both in the supersonic and the subsonic regions. The main result is that the transition region itself can be analyzed quite good with analytic EOS fit, and thus, for $\tau\geqslant0.25$ can be analysed with the extended Garnier's solver as well.

\subsection{Transition region in the critical case $\tau=\tau_c$ for \TaO foam}
\label{sec:Ta2O5_results}

Next, we analyze the transition region for \TaO. This material was chosen (asides gold) since it is used in many HEDP experiments, due to the fact that like gold it contains a high-Z element, which usually has a high opacity~\cite{BackPRL,Back2000}. Specifically, it is used as low-density (diluted) foam of order of $~0.1 \mathrm{g/cm^3}$. The analysis is done for the critical case ($\tau=\tau_c$) using the Garnier's-like solution, and using the analytic approximation for $\tau=\tau_c$ and also for $\tau=-0.05,0$ and $0.05$, since these are the exponents of the temperature-temporal power-law profile of the hohlraum experiment~\cite{exp_PRL} (see also next section). We do not use the extended-Garnier's-like solution in this region, since we have shown that it is not valid for $\tau\approx 0$.  

The analysis for this material requires a set of parameters for the self-similar model, as in the case of gold. The required parameters were obtained from fitting both EOS and OPACITY tables to the form of power-laws as in Eqs.~\ref{pwrlaws} at the thermodynamic region of $100-300\mathrm{eV}$. The tables we used are a QEOS EOS table (see~\cite{QEOS}), and a CRSTA table (see~\cite{Kurz2012, Kurz2013}) for the opacity. The values of the parameters obtained from both fits are specified in table~\ref{table:pwr_law_opac_eos_Au} (see also~\cite{avner1}). $\tau_c\approx 0.328$ for these fitted parameters.  The self-similar Garnier's-like solution for \TaO at $\tau_C$ and the resulting dimensionless parameters are shown in Appendix C.
\TaO has higher heat capacity and pressure compared to gold, and therefore higher speed of sound for a given temperature and density. Moreover, for densities lower than $1 \mathrm{g/cm^3}$ and temperatures lower than 100eV, \TaO has a higher opacity, as well, relatively to gold. That leads to a lower Mach number of \TaO, for a given temperature, density and time, compared to gold. Hence, the supersonic-subsonic transition occurs at a lower initial density for \TaO, than for gold.

\begin{figure}
		\includegraphics[width=7.5cm,clip=true]{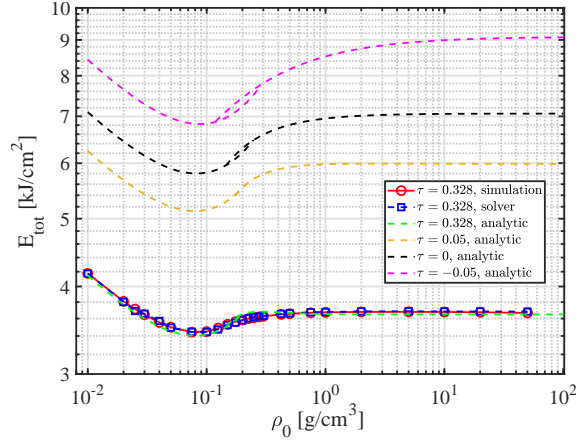}
		\caption{The energy per area unit Vs. the initial density $\rho_0$ for \TaO at $T_0=100eV$, $t=1nsec$ for various values of $\tau$, including the critical case ($\tau=\tau_c$) and for $\tau=-0.05,0$ and $0.05$, the fitted exponents for the temperature-temporal profile of the hohlraum experiment~\cite{exp_PRL}.}
		\label{fig:E_tot_vs_rho0_tau_c_Ta2O5_T0=100eV_t=1nsec}
\end{figure}

Similarly to gold, we examine the same physical quantities to characterize the transition region for \TaO for $T_0=100 \mathrm{eV}$ and $t=1 \mathrm{nsec}$. In Fig.~\ref{fig:E_tot_vs_rho0_tau_c_Ta2O5_T0=100eV_t=1nsec} we show the total energy per area unit as a function of the initial density for different $\tau$'s. For $\tau=\tau_c$ we compare Garnier's solution, with numerical simulations and the extended analytic approximation. Again, qualitatively the curves look much alike, as for gold. The transition region is obtained near the minimum point, for $\chi_0\approx1.2$, corresponds to $\rho_0=0.125 \mathrm{g/cm^3}$. This value is indeed lower than the one we have seen for gold, $\rho_0=0.2 \mathrm{g/cm^3}$), as expected.

The agreement between the solver and the simulations results is very good - better than $0.5\%$ for all values of $\rho_{0}$. The analytic approximation shows a relatively good agreement as well, better than $2\%$ to the simulations results, for all regions, except for the patching area of the approximated supersonic and subsonic branches (up to $5\%$). We can see that as a preparation for the comparison between the model and the experiment, the energy in the $\tau\approx 0$ region changes and is quite sensitive for the exact value of $\tau$. However, the minimum energy remains at $\rho_0\approx0.1 \mathrm{g/cm^3}$ for all $\tau$'s

\section{Comparison to Young's et al. experiment}
\label{sec:PRL_exp_results}

In this section the experimental results from~\cite{exp_PRL} are reproduced applying the analytic approximation using results for gold and \TaO introduced in the previous sections.
The main goal of the experiments was to examine whether the drive temperature in a hohlraum made from a low-density foam would be higher, due to less energy losses to the hohlraum walls. Three types of targets were fabricated for those experiments, all of them are cylindrical hohlraums (see Fig.~\ref{plots_0D}(c) in Appendix D). The first type is the standard gold hohlraum 
for a reference hohlraum, relatively to the other two. 
The second and third types are \TaO, one with initial density of $\rho_0=4\mathrm{g/cm^3}$ and a foam \TaO with $\rho_0=0.1\mathrm{g/cm^3}$. 

The experimental setup, including the geometrical dimensions of the target, is shown in Fig.~\ref{plots_0D}(c). 
The radiation temperature measurements are performed through a diagnostic hole. 
The \TaO hohlraums include a thin layer of gold at the left edge, so that for all targets, the laser beams hit the same material (gold), and the X-ray radiation emitted would have similar properties. For each shot 15 laser beams were used, each with an approximate energy of $\sim350\mathrm{J}$. This total amount of laser energy ($\sim5.25\mathrm{kJ}$) leads to drive temperatures of order of $\sim100\mathrm{eV}$.
After a short rise time ($\sim0.35\mathrm{nsec}$) 
the temperature for all cases is approximately constant, as the temporal profiles look quite flat. Nevertheless, comparing the temperature between the high and low densities for the \TaO hohlraums, a higher drive temperature is indeed achieved for the lower density. 

To use the self-similar expressions, we approximate the wall temperature in the experiment, 
which can be taken from the experimental results~\cite{exp_PRL}, as a power law in time. Since there is some uncertainty in the experimental results, 
we take an approximate value of $\tau=0$, but with some uncertainty of $\pm{0.05}$. These upper and lower bound on the value of $\tau$, would be used to examine the sensitivity of our results, to the input boundary condition. Another reason for this uncertainty is that the temporal profile of the three radiation temperatures $T_{\mathrm{Rad}}$, $T_W$, and $T_{\mathrm{obs}}$ (in a given spatial spot, see Appendix D), is different approximately by $\pm{0.05}$ in $\tau$ (see~\cite{alum}), when the measure is $T_{\mathrm{obs}}$, and the self-similar solution uses $T_W$.

Since the X-ray converting zone is always a gold zone (a primary ``P" hohlraum), then the X-rays flow to the secondary ``S" hohlraum, an extended zero-dimensional (0D) model is derived for calculating the drive temperature in specialized hohlraums. The model rests on two foundations: the energy balance equations for specialized hohlraums~\cite{rosenScale,mordy_refs,lindl1995,rosen1996}, and the three different definitions of radiation temperatures {\bf{in each zone}}~\cite{mordy_lec,MordiPoster}. The model is derived in details in Appendix D.


First, we calculate the geometrical dimensions of the areas needed for the 0D model equations, from the hohlraum's geometry described in~\cite{exp_PRL}. The hohlraum's diameter and length are $D=1.6 \mathrm{mm}$ and $\ell=2.6 \mathrm{mm}$, respectively. There are two holes we need to take into consideration: the laser entrance hole and the diagnostic hole (located at the other base of the cylinder). Their radii are: $R_{h1}=0.8 \mathrm{mm}$ and $R_{h2}=0.6 \mathrm{mm}$. 
As mentioned, in all cases the laser beams hit a thin layer of gold, with an approximated length of $\ell_1=0.35 \mathrm{mm}$ (the exact length is not specified in~\cite{exp_PRL}, we evaluated the length from the schematic experimental setup in Fig.~\ref{plots_0D}(c)). Thus, we have to take into account the area between the ``P" zone which is the always a gold layer, and the ``S" zone which is the rest of the hohlraum, made of \TaO or gold. We recalculate the areas for the two-zone 0D model: 
$A_{H,\mathrm{P}}=\pi{R_{h1}}^2=2.01 \mathrm{mm^2}$, $A_{H,\mathrm{S}}=\pi{R_{h2}}^2=1.13 \mathrm{mm^2}$, $A_{W,\mathrm{P}}=\pi\ell_1{D}+\pi\frac{D^2}{2}-\pi{R_{h1}}^2=1.76 \mathrm{mm^2}$, $A_{W,\mathrm{S}}=\pi(\ell-\ell_1){D}+\pi\frac{D^2}{2}-\pi{R_{h2}}^2=12.19 \mathrm{mm^2}$, and the area of the interface between the zones: $A_{\mathrm{PS}}=\pi\frac{D^2}{2}=2.01 \mathrm{mm^2}$. 

\begin{figure}[htp]
	\centering
	(a) \includegraphics[width=7.5cm,clip=true]{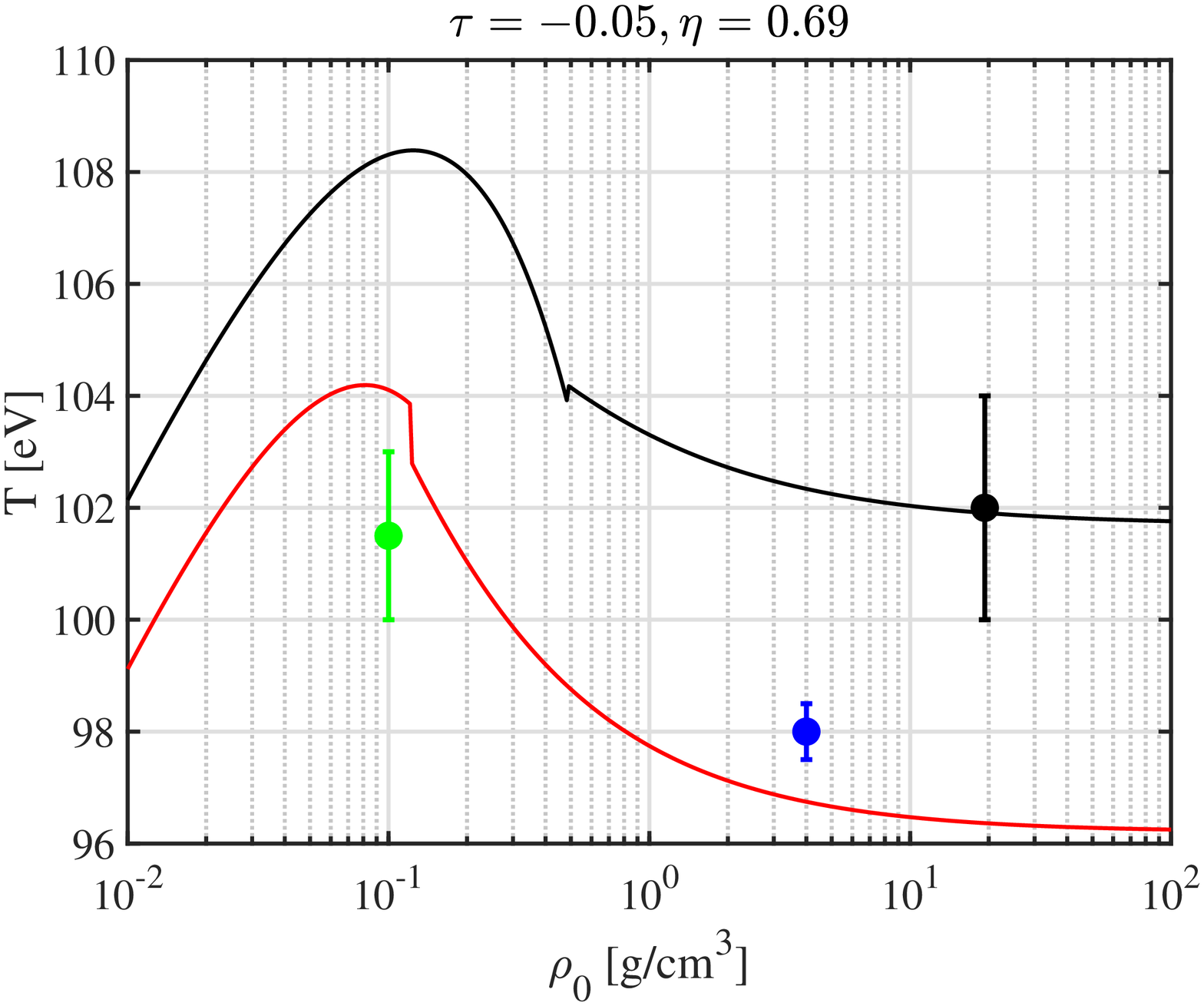}
	(b)
	\vspace{5mm}
	\includegraphics[width=7.5cm,clip=true]{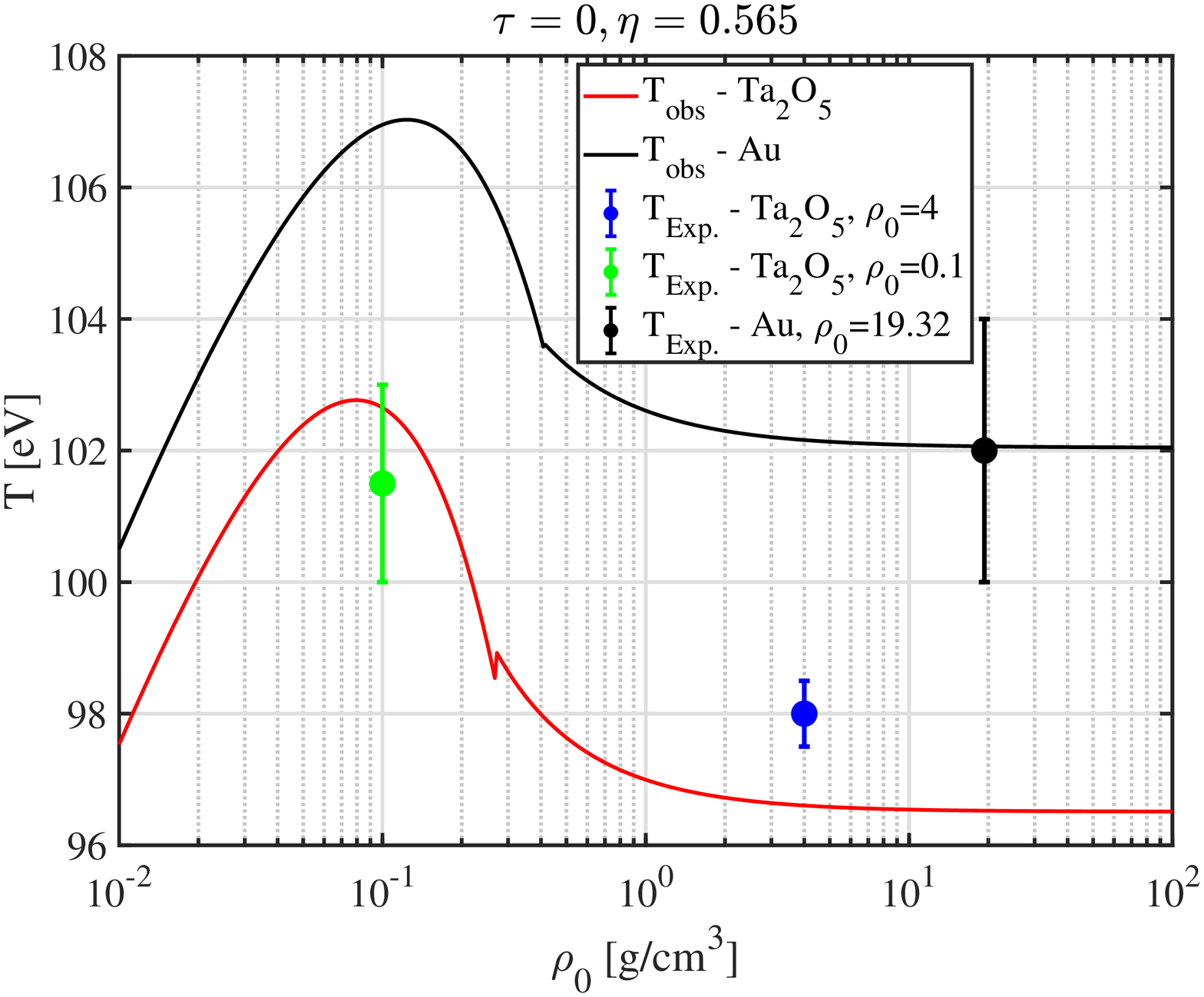}
	(c) \includegraphics[width=7.5cm,clip=true]{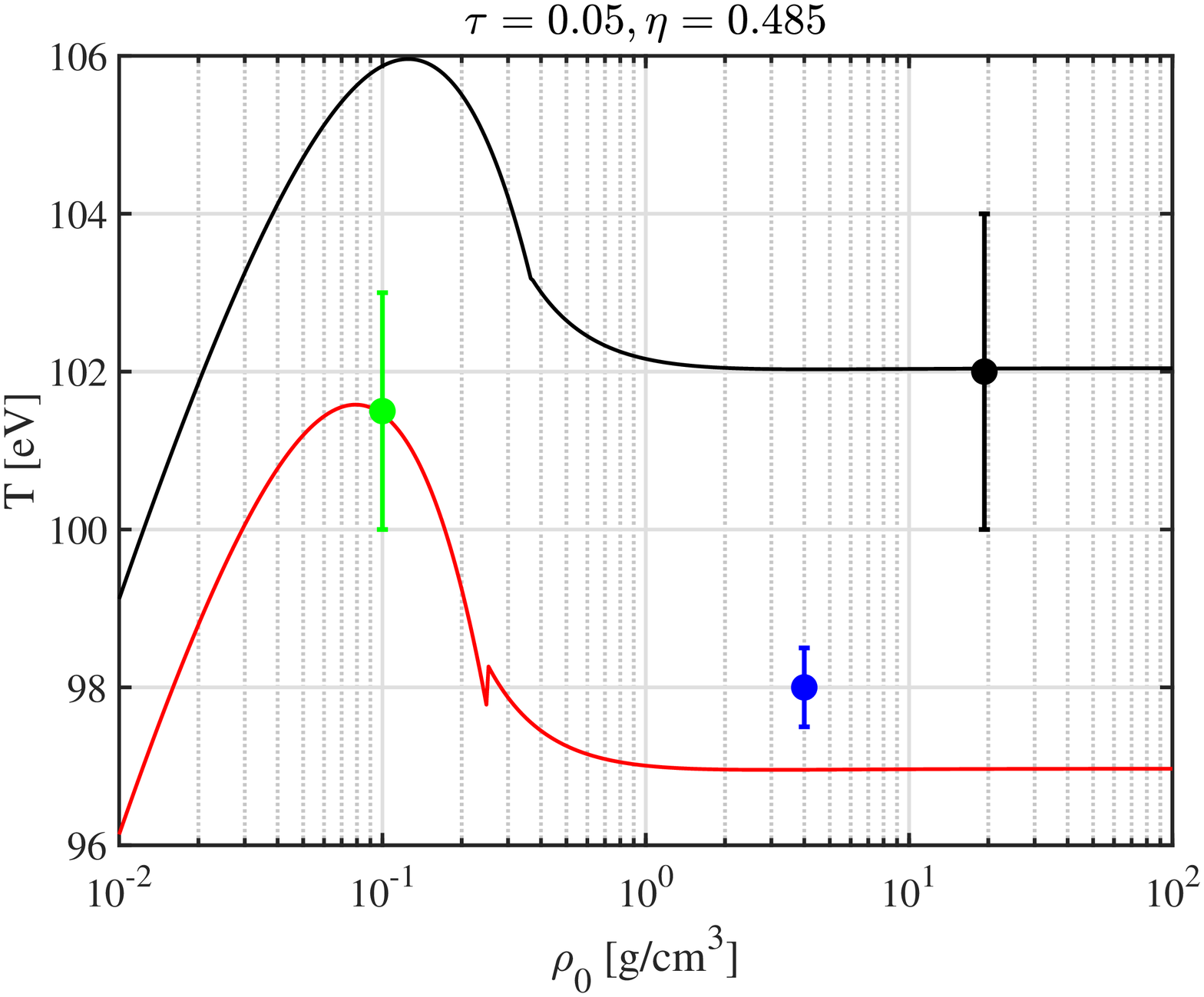}
	\caption{The radiation temperature $T_{\mathrm{obs}}$ as a function of the initial density $\rho_0$ for: (a) $\tau=-0.05$ and $\eta=0.69$, (b) $\tau=0$ and $\eta=0.565$, (c) $\tau=0.05$ and $\eta=0.485$. The two-zone 0D model results for gold are in the black curves and for \TaO in the red curves. The experimental results are shown for solid gold (black circle), solid \TaO (blue) and \TaO foam (green).}
	\label{fig:exp_results}
\end{figure}

Now using the model derived in Appendix D, Eqs.~\ref{0D_P_S} can be solved for the gold hohlraum, for a given value of the conversion efficiency coefficient $\eta$. Since we do not know the exact value of $\eta$, we use the result for the drive temperature in the {\em{gold hohlraum}} to calibrate the value of $\eta$, separately for each of the three values of $\tau$ (it is a reasonable assumption that the value of $\eta$ does not change dramatically replacing the ``S" hohlraum). 
Substituting $A_{W,\mathrm{P}}$, $A_{W,\mathrm{S}}$, $A_{H,\mathrm{P}}$, $A_{H,\mathrm{S}}$ and $A_{\mathrm{PS}}$ in Eqs.~\ref{0D_P_S}, we obtain two conjugated nonlinear equations for $T_{0,P}$ and $T_{0,S}$ for each pair of values of $\tau$ and $\eta$:
\begin{subequations} \label{0D_P_S_real1}
	\begin{align} \label{0D_P_real1}
	& \eta{E_L}=e_{0,\mathrm{Au}}T_{0,\mathrm{P}}^{P_{T,\mathrm{Au}}}t^{P_{t,\mathrm{Au}}}\left(A_{W,\mathrm{P}}+\frac{A_{H,\mathrm{P}}}{2}+\frac{A_{\mathrm{PS}}}{2}\right)+ \\  
	& \sigma\frac{A_{H,\mathrm{P}}}{1+4\tau}{T_{0,\mathrm{P}}^4}t^{1+4\tau}+ \sigma\frac{A_{\mathrm{PS}}}{1+4\tau}({T_{0,\mathrm{P}}^4}-{T_{0,\mathrm{S}}^4})t^{1+4\tau}-e_{0,\mathrm{Au}}T_{0,\mathrm{S}}^{P_{T,\mathrm{Au}}}t^{P_{t,\mathrm{Au}}}\frac{A_{\mathrm{PS}}}{2} \nonumber
	\end{align}
	\begin{align} \label{0D_S_real1}
	& e_{0,\mathrm{Au}}T_{0,\mathrm{S}}^{P_{T,\mathrm{Au}}}t^{P_{t,\mathrm{Au}}}\left(A_{W,\mathrm{S}}+\frac{A_{H,\mathrm{S}}}{2}+\frac{A_{\mathrm{PS}}}{2}\right)+\sigma\frac{A_{H,\mathrm{S}}}{1+4\tau}{T_{0,\mathrm{S}}^4}t^{1+4\tau}- \\
	& e_{0,\mathrm{Au}}T_{0,\mathrm{P}}^{P_{T,Au}}t^{P_{t,\mathrm{Au}}}\frac{A_{\mathrm{PS}}}{2}-\sigma\frac{A_{\mathrm{PS}}}{1+4\tau}({T_{0,\mathrm{P}}^4}-{T_{0,\mathrm{S}}^4})t^{1+4\tau}=0 \nonumber
	\end{align}
\end{subequations}
while the connection between $T_{\mathrm{obs,gold}}$ and $T_{0,\mathrm{S}}$ is defined explicitly in Appendix D. Eqs.~\ref{0D_P_S_real1} use the {\em{subsonic}} value of gold which matches for solid gold initial density (like in~\cite{rosenScale,mordy_refs,lindl1995,rosen1996}).
We solve Eqs.~\ref{0D_P_S} for $\tau=-0.05,0,0.05$ for $T_0$, finding the suitable values of $\eta$ which yields $T_{\mathrm{obs,gold}}=102 \mathrm{eV}$ (the experimental result gold hohlraum~\cite{exp_PRL}).
The three obtained values are: $\eta=0.69, 0.565, 0.485$ for $\tau=-0.05,0,0.05$, respectively. The dependency of the observed temperature for the gold hohlraum on its initial density is shown (black curves) in Fig.~\ref{fig:exp_results}(a-c) for $\tau=-0.05,0,0.05$. The calibrated values of $\eta$ fit by definition the $T_{\mathrm{obs,gold}}$ at $\rho_0=19.3 \mathrm{g/cm^3}$ with the experimental value in each of the three cases. 

To obtain {\em{all}} the curves in Fig.~\ref{fig:exp_results}(a-c) $T_{\mathrm{obs,gold}}$ or $T_{\mathrm{obs,Ta_2O_5}}$ as a function of the initial density, we use the supersonic approximation for the energy in the supersonic region for the low densities, and the subsonic approximation for the energy in the subsonic region for higher densities. The transition between both branches, is arbitrarily determined to be the point of intersection with the higher $\rho_{0}$ value between the two branches in the energy curve. 

Next, We use the calibrated values of $\eta$ to calculate the observed temperatures for the \TaO hohlraums (both densities), and compare between the results obtained from the 0D model to their corresponding experimental values, by solving these equations: 
\begin{subequations} \label{0D_P_S_real}
	\begin{align} \label{0D_P_real}
	& \eta{E_L}=e_{0,\mathrm{Au}}T_{0,\mathrm{P}}^{P_{T,\mathrm{Au}}}t^{P_{t,\mathrm{Au}}}\left(A_{W,\mathrm{P}}+\frac{A_{H,\mathrm{P}}}{2}+\frac{A_{\mathrm{PS}}}{2}\right)+ \\  
	& \sigma\frac{A_{H,\mathrm{P}}}{1+4\tau}{T_{0,\mathrm{P}}^4}t^{1+4\tau}+ \sigma\frac{A_{\mathrm{PS}}}{1+4\tau}({T_{0,\mathrm{P}}^4}-{T_{0,\mathrm{S}}^4})t^{1+4\tau}-e_{\mathrm{tot,Ta_2O_5}}\frac{A_{\mathrm{PS}}}{2} \nonumber
	\end{align}
	\begin{align} \label{0D_S_real}
	& e_{\mathrm{tot,Ta_2O_5}}\left(A_{W,\mathrm{S}}+\frac{A_{H,\mathrm{S}}}{2}+\frac{A_{\mathrm{PS}}}{2}\right)+\sigma\frac{A_{H,\mathrm{S}}}{1+4\tau}T_{0,\mathrm{S}}^4t^{1+4\tau}- \\
	& e_{0,}{T_{0,\mathrm{P}}}^{P_{T,\mathrm{Au}}}t^{P_{t,\mathrm{Au}}}\frac{A_{\mathrm{PS}}}{2}-\sigma\frac{A_{\mathrm{PS}}}{1+4\tau}(T_{0,\mathrm{P}}^4-T_{0,\mathrm{S}}^4)t^{1+4\tau}=0 \nonumber
	\end{align}
\end{subequations}

The calculated observed temperatures in the zone of \TaO (``S"), $T_{\mathrm{obs,Ta_2O_5}}$, are shown (in the red curves) in Figs.~\ref{fig:exp_results}(a-c) for $\tau=-0.05,0,0.05$. First, we can see that for both gold and \TaO, the observed temperatures for all values of $\tau$ are higher for lower initial densities compared the solid density, i.e., the supersonic region can yield higher temperatures than the subsonic region. 
This is consistent with the theory and our expectations, as well as with the experimental results. The obtained results for the observed temperature in \TaO are lower than for gold, for any initial density, due to the fact that \TaO has higher heat capacity. Hence for a given energy absorption within the walls of the hohlraum, the drive temperature for \TaO is lower relatively to gold. This is evident from both the 0D model results (for all the values of $\tau$), as well as from the experimental results. 

The experimental difference between the measured drive temperatures for low-dense foam \TaO hohlraum ($\rho_0=0.1 \mathrm{g/cm^3}$) and the solid \TaO hohlraum ($\rho_0=4 \mathrm{g/cm^3}$) is $\approx4$eV~\cite{exp_PRL}. The differences that our model predicts between these two specific densities from Fig.~\ref{fig:exp_results} are: 7eV, 6eV, and 4.5eV, for $\tau=-0.05,0,0.05$, respectively. We can see that for the nominal value $\tau=0$ (Fig.~\ref{fig:exp_results}(b)) $T_{\mathrm{obs}}$ for \TaO foam agrees with the experimental result of 101.5eV, and for solid \TaO it misses the experimental value of 98eV by $\approx 1$eV. For $\tau=-0.05$ (Fig.~\ref{fig:exp_results}(a)) the obtained $T_{\mathrm{obs}}$ misses the experimental value also by $\approx2$eV for the foam and $\approx 1$eV the solid density. For $\tau=0.05$ (Fig.~\ref{fig:exp_results}(c)) we obtain a very good agreement between $T_{\mathrm{obs}}$ to the experimental value for the foam density, and miss the experimental value for the solid density by $\approx 1$eV. considering this is a simple analytic model, an error of $\approx 1$eV, ($\approx 1\%$), is reasonable.  

\section{Summary}
\label{sec:Summary}

In this work, we have investigated the transition region between supersonic and subsonic radiative heat flow via self-similar solutions and numerical simulations. We characterized this region through an examination of several physical quantities.
The transition region was introduced for both gold, which is the main material that hohlraums are made from, and \TaO, which is a foam-wall material that is offered for future hohlraums experiments. Experiments that use low-density foams might be in the intermediate sonic zone, that gives the major motivation for this study.

We have extended the exact solution of Garnier et al.~\cite{Garnier}, which is exact only for the critical case ($\tau_c\approx0.435$ in gold), for other values of $\tau$ (a general power-law boundary condition). The comparison between the generalization of the exact solution of Garnier and the simulations shows that for early times, for $\tau>0.25$ the extended self-similar solution yields a good accuracy in the hydrodynamic profiles, as well as the total energy. For lower values of $\tau$, this approximated generalization yields poor agreement. Nevertheless, the accuracy of the extended solution increases with time, and for late times we yield a good agreement with the simulations even for lower values of $\tau$.

We have also extended the analytic approximation for the energy from a specific boundary condition ($\tau=0$) offered in~\cite{HR_PRE}, to a general $\tau$ for both supersonic, subsonic and the transition regions. The extended analytic approximation yields a very good agreement with the simulations results, up to 5\%, for a large variety of parameters (densities, temperatures, times and $\tau$'s). 

We extended the physical interpretation of Garnier's self-similar solution, and suggested a few physical quantities, which can be used to characterize the different regions in the problem, specifically, the transition region. We examined the total energy, the compression ratio of the density at the shock front, and the maximal pressure. 
It was shown that these physical quantities have a significant sensitivity in the transition region, and therefore they can be used to characterize the heat wave and its Mach number. Nevertheless, many of the parameters that characterizes the transition region show reduced sensitivity to $\tau$. These results encourage the use of Garnier's solution for studying the transition region.
	
The sensitivity of our characterization to the EOS was also examined, comparing between the analytic EOS to an exact SESAME EOS table. It was shown that in general, the supersonic and the transition regions are much less sensitive to the EOS than the subsonic region, due to the formation of a 
strong shock in this region. Examining the sensitivity between gold and \TaO, 
we found that the qualitative behavior of the heat wave is quite similar, while the quantitative differences of the examined physical quantities are about 10\%-20\%.
	
We tested the theoretical predictions against the hohlraum's temperature experiments of~\cite{exp_PRL}. 
We have reproduced the experimentally measured radiation temperatures inside hohlraums, made from different materials with varying densities up to 1\%. Specifically, we reproduce, using the 2-Temp ``0-d" model, the main experimental result that solid gold yields higher temperature than solid \TaO, while low-density \TaO foam yields higher temperature than the solid one.    

The models and approximations described and developed in this work enhance significantly our understanding regarding radiative heat flow, coupled to hydrodynamics. These analytic solutions may also be helpful in verification and validation of numerical codes and their accuracy. In general, it was shown that self-similar solutions are powerful analytic tools, that can be used to improve our physical understanding of the problem, of the transition of radiative heat waves from the supersonic to subsonic region, in theory, as well as in future experiments.

\appendix
\section{Henyey, Zel'dovich and Henyey-like profiles, for the supersonic Marshak wave}
\label{henyey}
Full self-similar analytic solutions, for the whole temperature profile for the supersonic case can be found for two specific cases, with specific boundary conditions (specific $\tau$):
\begin{itemize}
       \item The Henyey solution~\cite{HR}, when the heat front $x_f$ propagates linearly in time. This corresponds to $\tau=\frac{1}{4+\alpha-\beta}$.
       \item Zel’dovich self-similar solution~\cite{zeldovich} for energy impulse (i.e., constant total energy). This corresponds to a negative $\tau$, $\tau=-\frac{1}{4+\alpha+\beta}$.
\end{itemize}
In this appendix we introduce shortly these two solutions, and present an approximate-expansion for a general $\tau$, using a Henyey-like profile.

\begin{figure}[th]
	\centering
	\includegraphics[width=7.8cm]{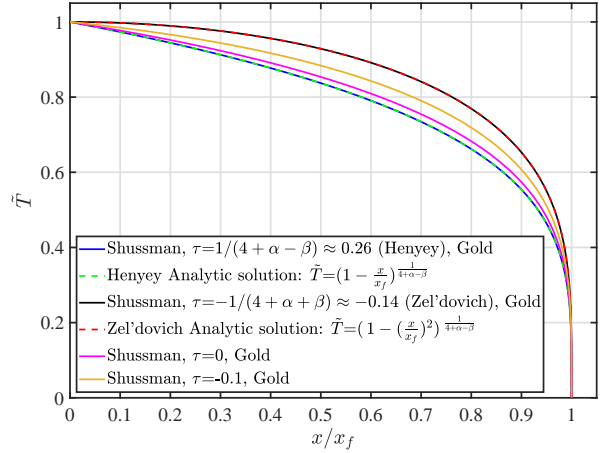}
	\caption{Self-similar temperature profile for several $\tau$ values. In addition, the analytic Henyey's and Zel'dovich's solutions (specific $\tau$'s) are presented.}
	\label{fig:self_similar_T_profiles}
\end{figure}

Henyey solved the supersonic Marshak wave problem in the case which the heat front is propagates linearly in $t$ (which corresponds to $\tau=\frac{1}{4+\alpha-\beta}$) already in 1954, but it was published much later (see the appendix in~\cite{HR}). The solution for the temperature profile in this case is in the form of $T(x,t)=T_s(t)\tilde{T}(\xi)$, where: $T_s(t)=T_0\left(\frac{t}{t_s}\right)^{\tau}=T_0\left(\frac{t}{t_s}\right)^{\frac{1}{4+\alpha-\beta}}$, and: $\tilde{T}(\xi)=(1-\xi)^{\frac{1}{4+\alpha-\beta}}$, where $\xi=x/x_f(t)$. $x_f(t)$ is the heat front position for this case is: $x_f(t)=\sqrt{\frac{16}{3\left(4+\alpha-\beta\right)}\frac{g\sigma{T_0}^{4+\alpha-\beta}}{f\rho_0^{2+\lambda-\mu}}}t$. For this case the heat front is linear in time, meaning the heat wave propagates at constant velocity. The dimensionless profile for Au parameters (Table~\ref{table:pwr_law_opac_eos_Au}) is shown by the green curve in Fig.~\ref{fig:self_similar_T_profiles}.


Another self-similar solution was introduced by Zel’dovich, for the case of a heat wave from an impulse instantaneous source (see~\cite{zeldovich,menahem}), for different geometries. Using~\citet{ps} notations, this problem is a constant total energy problem, which corresponds to $\tau=-\frac{1}{4+\alpha+\beta}$.
The problem was solved via a self-similar solution, based on dimensional analysis. The solution for the temperature is similar to the previous case: $T(x,t)=T_s(t)\tilde{T}(\xi)$. Here, $T_s(t)=T_0\left(\frac{t}{t_s}\right)^{\tau}=T_0\left(\frac{t}{t_s}\right)^{-\frac{1}{4+\alpha+\beta}}$, where $T_0=\left(\frac{3(4+\alpha)}{16}\frac{\rho_0^{\mu+\lambda}Q^2}{\sigma{gf}}\right)^{\frac{1}{4+\alpha+\beta}}$ and $Q$ denotes the initial energy per unit area within the source. The spatial profile $\tilde{T}(\xi)$ in this case is: $\tilde{T}(\xi)=\left(1-\xi^2\right)^{\frac{1}{4+\alpha-\beta}}$ (see the red curve in Fig.~\ref{fig:self_similar_T_profiles}). The dimensionless profile for Au parameters is shown by the red curve in Fig.~\ref{fig:self_similar_T_profiles}. In this case, the derivative of $d \tilde{T}/d\xi\to 0$ for $\xi\to 0$, due to the spatial dependency of $\tilde{T}(\xi)$.
For this case $\tau<0$, which implies the temperature at the origin decreases in time. This is actually expected, because the heat front propagates in time, as the total energy it contains stays constant.

Since the self-similar solution for the supersonic case in~\cite{shussman} is for a general power-law temperature profile (general $\tau$), the self-similar solutions of Henyey and Zel'dovich are private cases. Indeed, looking at the general form of the solution from~\cite{shussman} for the mass behind the heat front: $m_f(t)\propto{t^{\frac{1+\left(4+\alpha-\beta\right)\tau}{2}}}$, 
and for the total energy (per area unit): $E_{\mathrm{tot}}(t)\propto{t^{\frac{1+\left(4+\alpha+\beta\right)\tau}{2}}}$, one can easily see that a linear propagation of the heat front in time requires $\tau=\frac{1}{4+\alpha-\beta}$, like in Henyey's solution, and that conservation of the total energy yields $\tau=-\frac{1}{4+\alpha+\beta}$, like in the self-similar solution of Zel'dovich.  

In Fig.~\ref{fig:self_similar_T_profiles} we present several dimensionless profiles using different $\tau$'s of the exact self-similar solution for the supersonic case for Au~\cite{shussman}.
For the values of $\frac{1}{4+\alpha-\beta}\approx0.26$ (Henyey) and $-\frac{1}{4+\alpha+\beta}\approx-0.14$ (Zel'dovich) the self-similar solution reproduces the analytic solutions of Henyey and Zel'dovich, exactly. Moreover, noticing the two intermediate values - $0$ (for constant boundary temperature) and $-0.1$, yields an interesting observation. The dimensionless $\tau=0$ profile is much closer to Henyey profile (although with a different $x_f$), than Zel'dovich profile, despite the quite different value of $\tau$. Even for $\tau=-0.1$ (which is close to $\tau=-0.14$ of Zel'dovich solution), the profiles lies between the two profiles. Thus, for a general power-law boundary condition, Henyey's self-similar solution may serve as a better approximation than the self-similar solution of Zel'dovich, for the relevant range of $\tau$'s that involves hohlraums experiments.
We define a {\em{Henyey-like}} profile, taking $T(x,t)=T_s(t)\tilde{T}(\xi)$, where $T_s(t)=T_0\left(\frac{t}{t_s}\right)^{\tau}$ (using the general $\tau$) and $\tilde{T}(\xi)=(1-\xi)^{\frac{1}{4+\alpha-\beta}}$ taking $x_f(t)$ for the specific $\tau$ using the self-similar solution~\cite{shussman} or approximate solution~\cite{HR}. Moreover, $x_f(t)$ can be taken even numerically or from any {\em{external}} modeling, i.e., given a general $x_f(t)$ using the the Henyey function, yields pretty good approximation for whole the temperature profile. In~\cite{avner2,avner1} the Henyey-like profiles are specifically used, for estimating the heat front position $x_f(t)$ comparing to experiments, and for evaluating the energy leakage in supersonic Marshak waves experiments in low-density foams toward the opaque Au coating.


\section{Comparison of the approximated Shussman \& Heizler's solution to Garnier's exact solution for the critical case}
\label{Shussman_Heizler_vs_Garnier_profiles}

In this appendix we set an interesting comparison between the exact solution of Garnier et al. in the critical case of gold with $\tau_c=\frac{1}{4+\alpha-2\beta}\approx 0.43$~\cite{Garnier}, and the approximated Shussman and Heizler solutions~\cite{shussman}. The Garnier et al. solution is valid for both supersonic, intermediate and subsonic regimes, controlled via the dimensionless parameter $\chi_0$. 
Shussman and Heizler offers a supersonic solution (which corresponds to high $\chi_0\gg 1$) for the temperature profile (when hydrodynamic motion is negligible) and for the strong shock subsonic limit profiles (for a general $\tau$), by patching two self-similar solutions (which corresponds to $\chi_0\ll 1$). Both solutions are also compared to full numerical simulations.

\begin{figure}[htp]
	\centering
	\includegraphics[width=16cm]{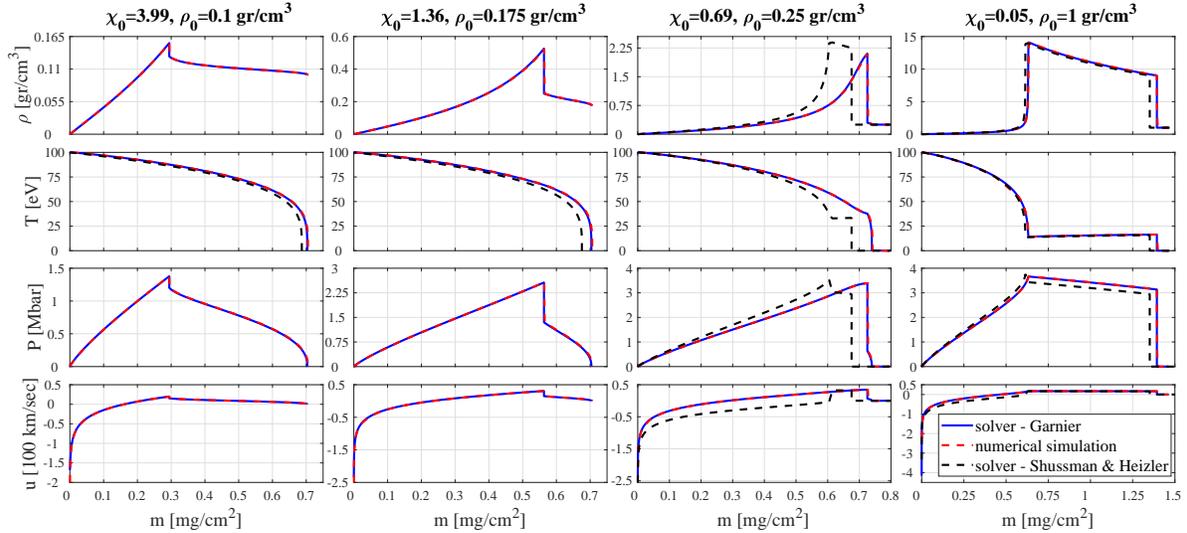}
	\caption{The hydrodynamic spatial profiles for various initial densities $\rho_0$ (and $\chi_0$) for gold at $T_0=100$eV, $t=1$nsec in the critical case ($\tau_c\approx 0.43$) using Garnier et al. exact self-similar solution (solid blue), the approximated Shussman and Heizler solution (dashed black) and numerical simulations (dashed red).}
	\label{fig:profiles_Au_tau_c_T0=100_t=1}
\end{figure}

The hydrodynamic profiles, the density, the temperature, the pressure and the velocity profiles as a function of the Lagrangian coordinate $m$, for several values of $\chi_0$ are shown in Fig.~\ref{fig:profiles_Au_tau_c_T0=100_t=1}. The chosen values are $\chi_0=3.99, 1.36, 0.69, 0.05$, for demonstrating the transition between the supersonic region to the subsonic regime.
These values correspond to initial densities of $\rho_0=0.1, 0.175, 0.25, 1 \mathrm{g/cm^3}$, for $T_0=100 \mathrm{eV}$ and $t=1 \mathrm{nsec}$.

First, Garnier's solution matches the numerical simulations perfectly, for all values of $\chi_0$. Next, for $\chi_0=3.99$, the heat wave is supersonic, when the hydrodynamics motion is small and thus, the pure supersonic solution of Shussman matches the temperature profile. For $\chi_0=1.36$, the wave is still supersonic, thus the temperature profiles still yields a good approximation, however, it is close to the sonic point so hydrodynamics starts playing a significant roll. For $\chi_0=0.69$, the heat wave is already subsonic, and the shock wave can be seen clearly in the density as well as in the pressure profiles. 
Finally, for $\chi_0=0.05$ the heat wave is in the deep subsonic regime. 
In this strong shock subsonic case, the agreement between the exact Garnier et al. solution and the Shussman and Heizler's approximated solution is very good, as expected. We do mention the possible advantage of the patching technique, which enables using a binary-equation of state that might be essential for modeling some real solid materials experiments~\cite{binary_EOS,alum}.


\section{Solution for the critical case for \TaO}
\label{Ta2O5_tau_c}

\begin{figure}[htp]
	\centering
	(a)
		\includegraphics[width=7.7cm,clip=true]{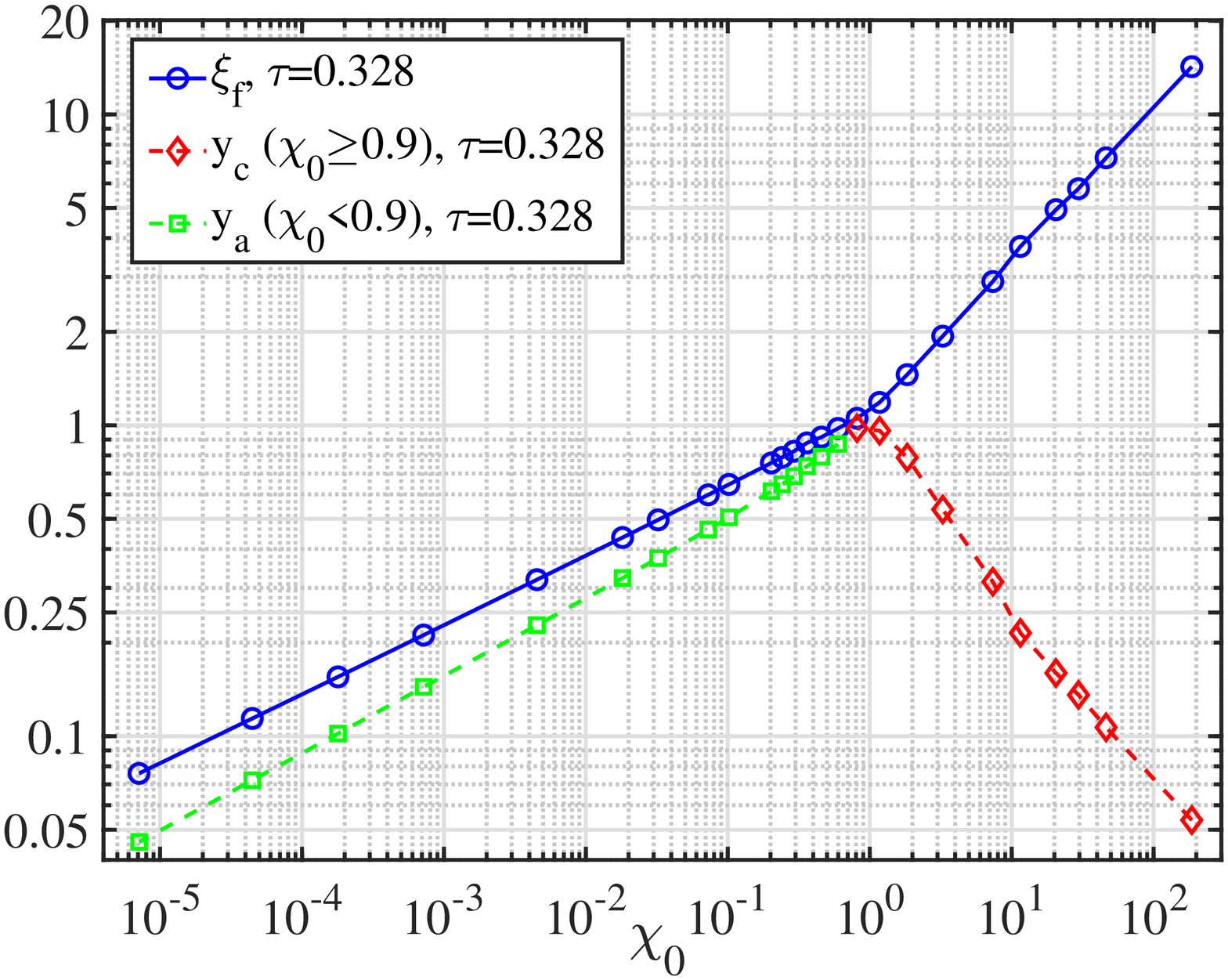}
    (b)
		\includegraphics[width=7.4cm,clip=true]{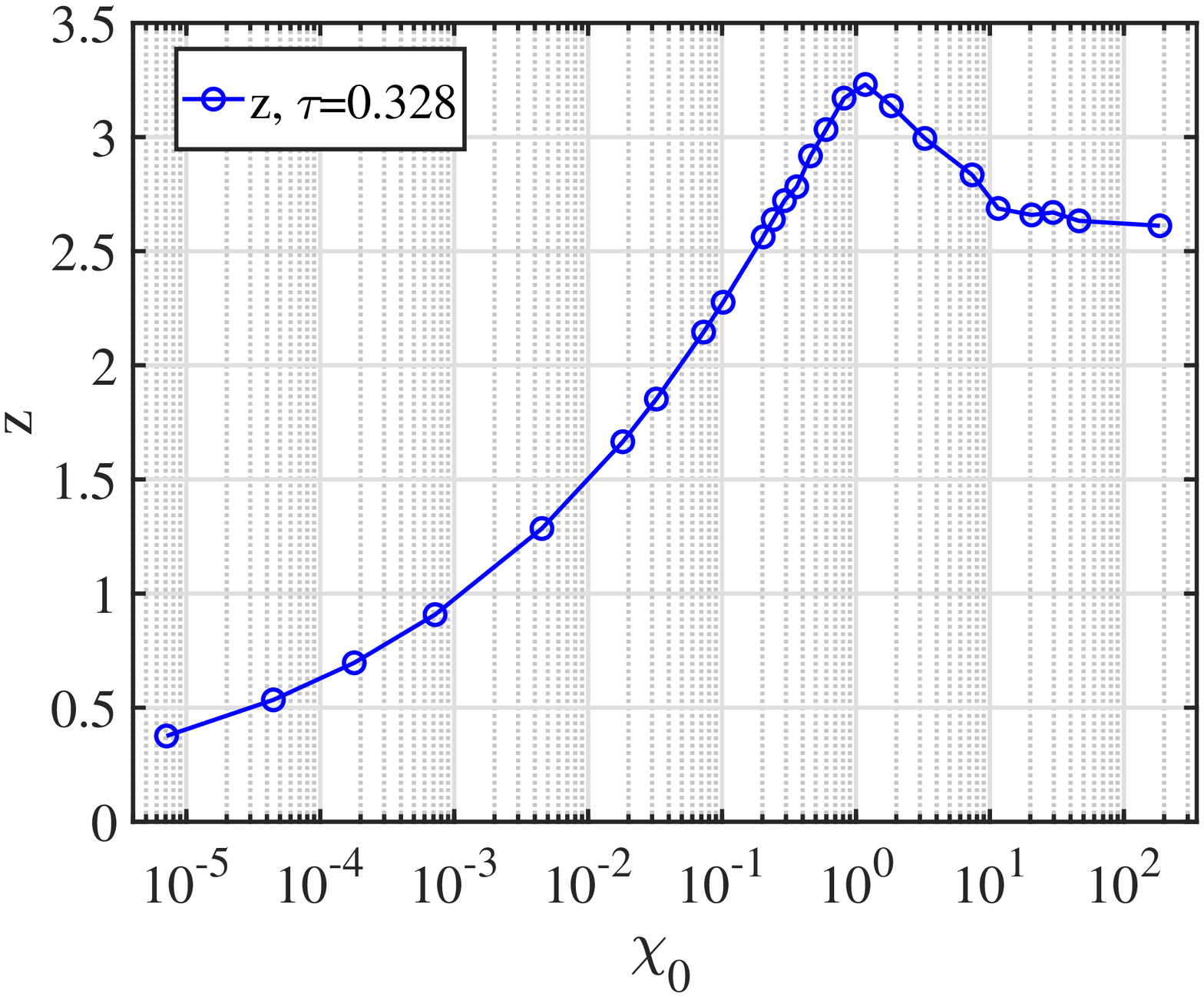}
		\caption{(a) The dimensionless parameters of the self-similar solution for \TaO as a function of $\chi_0$ in the critical case ($\tau=\tau_c$). (b) The energy dimensionless parameter of the self-similar solution $\ti{z}$ for \TaO as a function of $\chi_0$ in the critical case ($\tau=\tau_c$).}
		\label{fig:xsi_f_y_and_z_tilda_vs_chi0_SOLVER_for_Ta2O5_tau_c}
\end{figure}

In this appendix, we apply the self-similar solver for the critical case, for $\tau_c=0.328$ for \TaO, and find the dimensionless self-similar parameters, using the analytic parameters for the EOS and the opacity. The parameters $\xi_f$, $y_c$ and $y_a$ are presented as a function of $\chi_0$ in Fig.~\ref{fig:xsi_f_y_and_z_tilda_vs_chi0_SOLVER_for_Ta2O5_tau_c}(a). The dimensionless parameter of the energy 
$\ti{z}$ is presented as a function of $\chi_0$ in Fig.~\ref{fig:xsi_f_y_and_z_tilda_vs_chi0_SOLVER_for_Ta2O5_tau_c}(b). For both sets of parameters we can see qualitatively the same behavior as we have seen for gold. $\xi_f$ is increasing with $\chi_0$, and changes its slope at the transition region (near $\chi_0=0.9$ for \TaO). $y_c$ increases with $\chi_0$ to $\chi_0=0.9$, and for larger values of $\chi_0$ the solver yields $y_a$ values instead, since it is in the strong shock regime. Here $y_a$ decreases with $\chi_0$, as before.

The dimensionless energies also show the same type of behavior as for gold. $\ti{z}$ has a maximum at the transition region for $\chi_0\approx1$. Most of the energy for all values of $\chi_0$ is internal, but the fraction of kinetic energy from the total energy increases as $\chi_0$ decrease. It is about $~15\%$ from the total energy at the transition region for $\chi_0=1$.  

\section{Primary-Secondary (2-Temp) ``0-dimensional" hohlraum temperature model}
\label{0dim}

In this appendix, a zero-dimensional (0D) model for calculating the drive temperature in specialized hohlraums is presented. The model is rest on two foundations, both of them of M.D. Rosen: The first one is the energy balance equations for specialized hohlraums (resulting spatially two-temperatures hohlraum model, i.e. a ``primary" (``P") hohlraum  and a ``secondary" (``S") hohlraum temperatures)~\cite{rosenScale,mordy_refs,lindl1995,rosen1996}. The second one is the important difference between three different radiation temperatures {\bf{in each hohlraum}} (primary and secondary)~\cite{mordy_lec,MordiPoster}: The drive temperature $T_{\mathrm{Rad}}$, which is the temperature that characterized the incident flux toward the hohlraum's wall, the wall surface temperature $T_W$, and the temperature of the emitted flux $T_{\mathrm{obs}}$, that an x-ray detector would measure, which is approximately the temperature for 1mfp inside the sample (see also in~\cite{avner1,avner2,alum}). The model is called ``0d" since there is no {\em{spatial}} temperature dependency inside each hohlraum (besides the ``P"-``S" deviation).  Since in the experiment of Young et al., the laser is shot to an ``outer" gold hohlraum, where the laser energy is converted to X-ray~\cite{exp_PRL}, this model is important in reproducing the experimental results.

\begin{figure}[th]
	\centering{
		(a)
		\includegraphics[width=6.3cm]{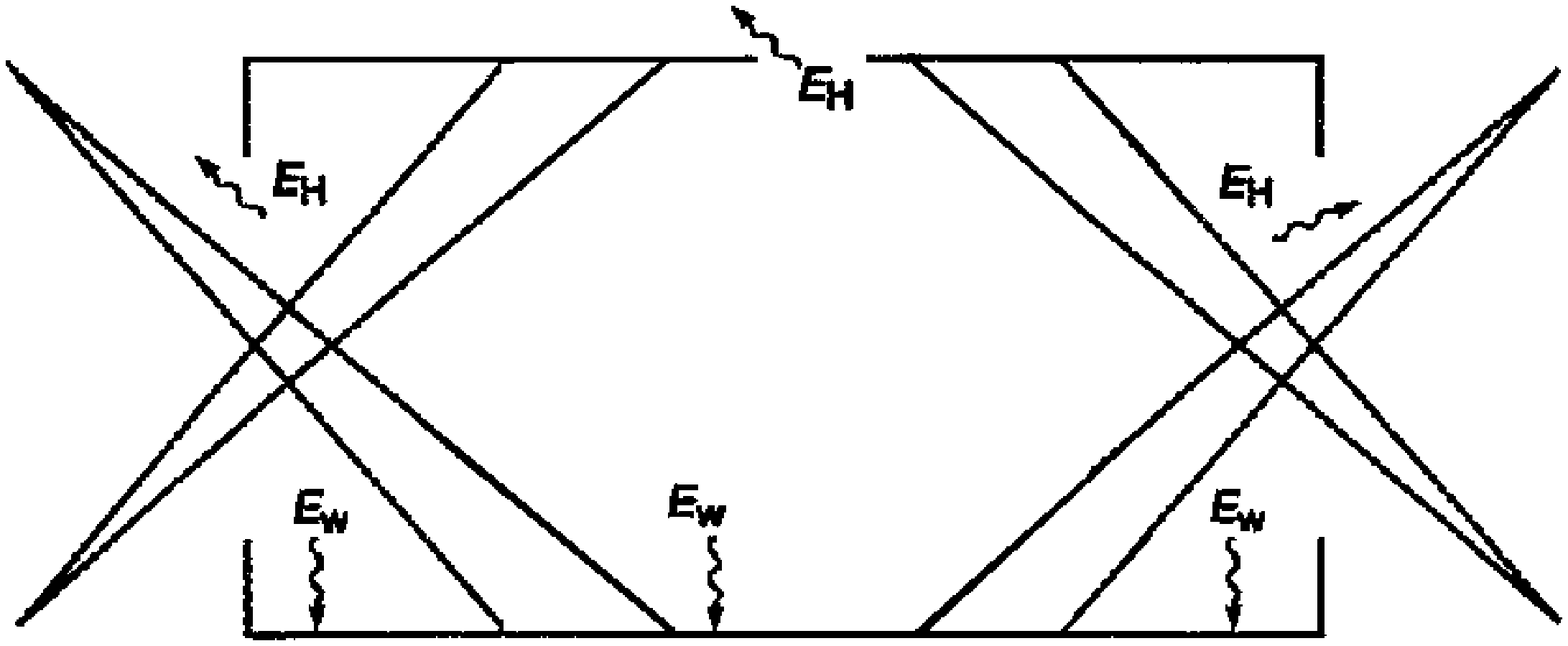}
		(b)
		\includegraphics[width=6cm]{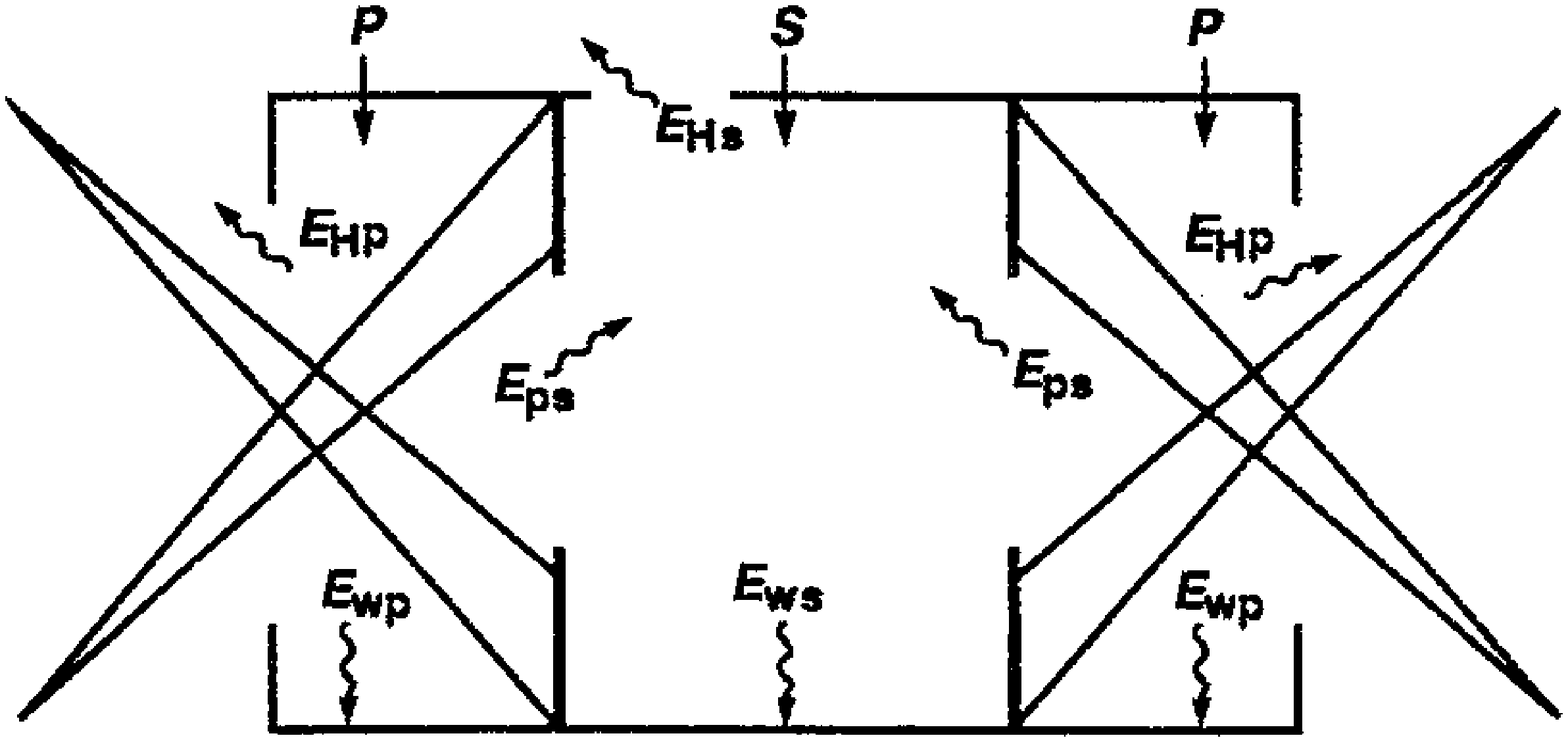}
		\\(c)
		\includegraphics[width=10cm]{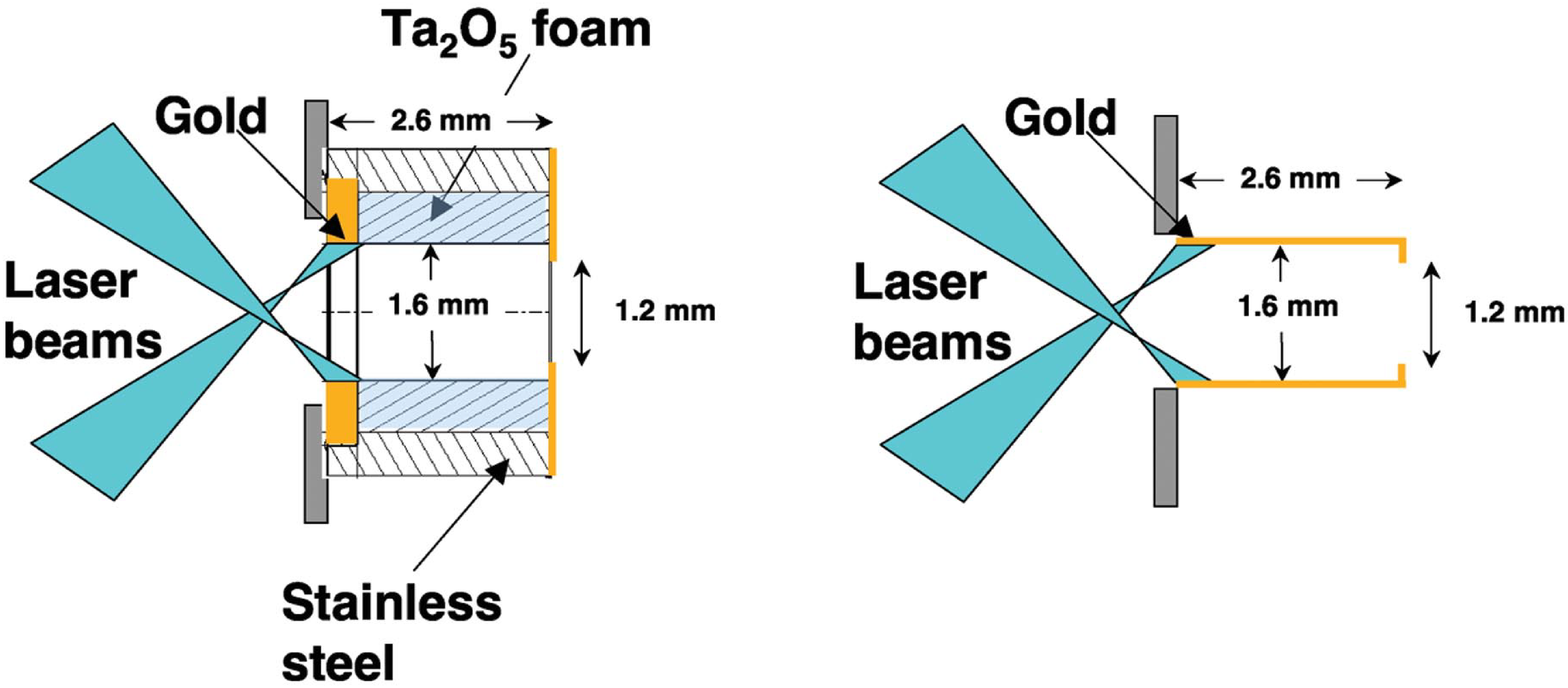}
	}
	\caption{A schematic hohlraum with the energy terms for the 0D model for: (a) a single region, (b) two regions (p-primary, s-secondary). (c) The Young et al. experimental setup, when in the left side, a \TaO hohlraum target, and in the right side, a gold hohlraum target.
	(a-b) are reproduced with permission from M.D. Rosen, Phys. Plas.~3, 1803 (1996)~\cite{rosen1996}. Copyright 1996 AIP Publishing. (c) is reproduced with permission from P.E. Young et al., \prl~101, 035001 (2008)~\cite{exp_PRL}. Copyright 2008 American Physical Society.}
	\label{plots_0D}
\end{figure}

The 0D model was first proposed by M.D. Rosen for modeling simple hohlraums (see~\cite{rosenScale,mordy_refs,lindl1995,rosen1996,mordy_lec}), and is based on a balance of energy within the hohlraum, between the incoming laser beams energy, the energy emitted outside the hohlraum through its holes and the energy lost by absorption into the hohlraum's walls. A schematic view of the system is presented in Fig.~\ref{plots_0D}(a). Given the hohlraum geometrical dimensions and the laser power as a function of time $P(t)$, the time-dependent drive temperature can be calculated. Thus, the basic equation which describes the energy balance within the hohlraum is:
\begin{equation}
\label{0D_energy_balance} 
\eta{E_L}(t)=E_W(t)+E_H(t)
\end{equation}
where $\eta$ is the conversion efficiency of laser energy to X-ray energy, when a typical value for gold is usually 50-75\%~\cite{kauff,mordy_refs,rosen1996}. $E_L(t)$ is the time-dependent integrated laser energy, entering into the hohlraum,  $E_L(t)=\int_{0}^{t}{P(t')dt'}$. $E_W$ is the energy absorbed in the hohlraums walls and $E_H$ is the energy leaking outside the hohlraums holes. 

Usually, $E_W$ is calculated from the self-similar {\em{subsonic}} solution for a heat wave in the material the wall is made of (usually a solid  gold)~\cite{ps,ps2,ger3,rosenScale,rosenScale2,rosenScale3,rosen1996,HR,Garnier,shussman}. Therefore, the wall energy per unit area has the form of:
\begin{equation}
\label{0D_wall_energy} 
E_{W}(t)=A_WE_{0}{T_0}^{p_{T}}{\rho_0}^{p_{\rho}} {t}^{p_{t}},
\end{equation}
where $A_W$ is the hohlraum wall area. In the case of low-opacity material or low-density foams (where the solution is not in the deep subsonic regime), Eq.~\ref{0D_wall_energy} will be replaced of course with the more general corrections that we have derived previously, Eq.~\ref{final_super_cor} or Eq.~\ref{final_sub_cor}, for supersonic and subsonic cases, respectively. 
The leaking energy through the holes $E_H$ is calculated via an integration in time over the outcoming flux through the hole (i.e. the incident flux to a unit area, wall or hole) $\sigma T^4_{\mathrm{Rad}}$. From Marshak boundary condition (see~\cite{mordy_lec,MordiPoster,avner1,avner2}), which is obtained by integration of the angular dependency of the flux for all incoming angles, it holds that:
\begin{equation}
\label{0D_Marshak_BC} 
\sigma T^4_{\mathrm{Rad}}=\sigma T^4_{W}+\frac{1}{2}\frac{\dot{E}_{W}(t)}{A_W} 
\end{equation}
where $\sigma T_W^4$ corresponds to the flux of the wall temperature (recalling that the wall temperature is given by $T_W=T_0{t}^\tau$). From this relation we deduce that the radiation temperature $T_{\mathrm{Rad}}(t)$ is {\em{always higher}} than the wall temperature $T_W(t)$. Thus the total energy that leaked through the hole is carrying out via the integral in time (defining $A_H$ as the hole's area):
\begin{align}
\label{0D_P_for_T0_1}
& E_{H}(t)=A_H\int_{0}^{t}{\sigma T^4_{\mathrm{Rad}}(t')dt'}=A_H\int_{0}^{t}{\left(\sigma T^4(t')_W+\frac{1}{2}\dot{E}_{W}(t')/A_W\right)dt'}= \\
& A_H\int_{0}^{t}{(\sigma{\left(T_0{t'}^\tau\right)}^4)dt'}+\frac{A_H}{2}\frac{E_{W}(t)}{A_W}=A_H\frac{\sigma{T_0}^4{t}^{1+4\tau}}{1+4\tau}+\frac{A_H}{2}\frac{E_{W}(t)}{A_W}. \nonumber
\end{align}
where $A_H$ is the total hohlraum holes area.

Finally, we substitute the expressions for 
$E_W(t)$ and $E_H(t)$ in Eq.~\ref{0D_energy_balance}, yielding a transcendental but closed equation for $T_0(t)$ for any given time (and thus, for $T_W(t)$ by $T_W(t)=T_0{t}^\tau$)~\cite{MordiPoster,Thomas2008}:
\begin{equation}
\label{0D_P_for_T0_2}
\left(A_W+\frac{A_H}{2}\right)\frac{E_{W}(t)}{A_W}+A_H\frac{\sigma{T_0}^4{t}^{1+4\tau}}{1+4\tau}-\eta E_L(t)=0.
\end{equation}
The radiation temperature $T_{\mathrm{Rad}}(t)$ is found later from Eq.~\ref{0D_Marshak_BC}. 

Lastly, Finding the third radiation temperature, the temperature which is measured from an external diagnostic, $T_{\mathrm{obs}}(t)$, we use a locally energy conservation, 
$\sigma{T_{\mathrm{Rad}}}^4(t)=\sigma T_{\mathrm{obs}}^4(t)+\dot{E}_{W}/A_W$. Using Eq.~\ref{0D_Marshak_BC} again yields, $\sigma T_{\mathrm{obs}}^4(t)=\sigma{T_{\mathrm{Rad}}}^4(t)-\dot{E}_{W}/A_W=\sigma{T_W}^4(t)-\frac{1}{2}\dot{E}_{W}/A_W$.
From this relation, we deduce that the observed temperature $T_{\mathrm{obs}}(t)$, is {\em{always lower}} than the wall temperature $T_W(t)$.

The described 0D model enables evaluation of the different radiation temperatures for a simple single hohlraum. However, in more complex-geometry hohlraums, where there is a large spatial deviation in the radiation temperatures, this simple model yields only partial averaged results. For example, the schematic hohlraum described in Fig.~\ref{plots_0D}(b) contains two distinct zones, labeled as ``P" (for primary) and ``S" (for secondary). Rosen expanded his simple model for complex hohlraums of these kinds~\cite{rosenScale,mordy_refs,lindl1995,rosen1996}. In each zone, the simple model can be applied, and we need to account for the interaction and energy exchanges between near zones. The incoming laser energy is absorbed only in the ``P" zone, some of the converted X-ray energy leaks from zone ``P" to zone ``S" (and vise-verse). Since in general there is no symmetry between the zones, each zone is characterized by a different set of temperatures $T_W,T_{\mathrm{Rad}},T_{\mathrm{obs}}$.
This complex hohlraum can be described by two equations, similarly to Eq.~\ref{0D_energy_balance}:
\begin{subequations} \label{0D_P_S}
	\begin{equation} \label{0D_P}
	\eta{E_L}(t)=E_{W,\mathrm{P}}(t)+E_{H,\mathrm{P}}(t)+E_{\mathrm{PS}}(t)
	\end{equation}
	\begin{equation} \label{0D_S}
	E_{\mathrm{PS}}(t)=E_{W,\mathrm{S}}(t)+E_{H,\mathrm{S}}(t)
	\end{equation}
\end{subequations}  

Eq.~\ref{0D_P} represents the energy balance in zone ``P", and counts for the energy exchange with zone ``S" as an energy {\em{net}} loss (loss - gain) ($E_{\mathrm{PS}}(t)$), in addition to the losses due to the walls absorption and the holes in this zone ($E_{W,\mathrm{P}}(t)$ and $E_{H,\mathrm{P}}(t)$). It treats the interface area $A_{\mathrm{PS}}$ as a hole which energy leaks from ``P" to``S" and vise-verse, each zone with its $T_{\mathrm{Rad}}$:
\begin{align} \label{energy_PS}
	& E_{\mathrm{PS}}=A_{\mathrm{PS}}\int_{0}^{t}{\left(\sigma T^4_{\mathrm{Rad,P}}(t')-\sigma T^4_{\mathrm{Rad,S}}(t')\right)dt'}=\\
	& \frac{\sigma A_{\mathrm{PS}}t^{1+4\tau}}{1+4\tau}\left({T_{0,\mathrm{P}}^4}-{T_{0,\mathrm{S}}^4}\right)+\frac{A_{\mathrm{PS}}}{2}\left(\frac{E_{W,\mathrm{P}}(t)}{A_{W,\mathrm{P}}}-\frac{E_{W,\mathrm{S}}(t)}{A_{W,\mathrm{S}}}\right) \nonumber
\end{align}
On the other hand, $E_{\mathrm{PS}}(t)$ serves as the {\em{net}} energy source for zone ``S", where the sinks of energy for this zone are $E_{W,\mathrm{S}}(t)$ and $E_{H,\mathrm{S}}(t)$. 

Using this 2-Temp model, Rosen has shown a good agreement between the measured temperature in various complex hohlraums experiments, like the hohlraum in Fig.~\ref{plots_0D}(b)~\cite{rosenScale,mordy_refs,lindl1995,rosen1996}.
We are interested in analysing the hohlraum experiment~\cite{exp_PRL} showed in Fig.~\ref{plots_0D}(c). We can see that the laser beams are shot into an outer gold zone, and from there the X-ray moves toward the major part of the hohlraum ($\mathrm{Ta_2O_5}$ or Au). Analysing the pure gold hohlraum by the simple one-zone hohlraum model Eq.~\ref{0D_energy_balance} yields a very low value of $\eta$, which is very far from the past experiments. Using the 2-Temp model Eqs.~\ref{0D_P_S} yields values of 50-70\%, which are comparable to the past experiments (see Fig.~\ref{fig:exp_results}). 

\begin{acknowledgments}
We acknowledge the support of the PAZY Foundation under Grant \textnumero~61139927.

\section*{Data Availability}
The data that support the findings of this study are available from the corresponding author upon reasonable request.
\end{acknowledgments}

\bibliography{bibliography}
\end{document}